\documentclass{aa}
\usepackage{geometry}
\usepackage{physics}
\usepackage{graphicx}
\usepackage{textcomp}
\usepackage{amssymb}
\usepackage{xcolor}
\usepackage{gensymb}
\usepackage{hyperref}
\usepackage{url}
\usepackage{multirow}
\usepackage{booktabs}
\usepackage{multirow}
\usepackage{pdflscape}

\newcommand{\valentin}{\textcolor{red}}

\begin{document}

\title{Fast radio burst repeaters produced via Kozai-Lidov feeding\\ of neutron stars in binary systems}

   \subtitle{}

   \author{V. Decoene
          \inst{1}\thanks{\email{decoene@iap.fr}}
          \and
          K. Kotera \inst{1}
          \and
          J. Silk \inst{1}\fnmsep \inst{2}\fnmsep \inst{3}
          }

   \institute{
   		Sorbonne Universit\'e, CNRS, UMR 7095, Institut d'Astrophysique de Paris, 98 bis boulevard Arago, 75014 Paris, France
	\and
		Department of Physics and Astronomy, The Johns Hopkins University, Homewood Campus, Baltimore, MD 21218, USA
	\and
		Department of Physics, University of Oxford, Keble Road, Oxford OX1 3RH, UK
             }

   \date{Received July 20, 2020; accepted November 26, 2020}
      
    \abstract{
    Neutron stars are likely surrounded by gas, debris, and asteroid belts. Kozai-Lidov perturbations, induced by a distant, but gravitationally bound companion, can trigger the infall of such orbiting bodies onto a central compact object. These effects could  lead to the emission of fast radio bursts (FRBs),  for example by asteroid-induced magnetic wake fields in the wind of the compact object. A few percent of binary neutron star systems in the Universe, such as neutron star-main sequence star, neutron star-white dwarf, double neutron star, and neutron star-black hole systems, can account for the observed non-repeating FRB rates. More remarkably, we find that wide and close companion orbits lead to non-repeating and repeating sources, respectively, and they allow for one to compute a ratio between repeating and non-repeating sources of a few percent, which is in close agreement with the observations.

    Three major predictions can be made from our scenario, which can be tested in the coming years: 1) most repeaters should stop repeating after a period between 10 years to a few decades, as their asteroid belts become depleted; 2) some non-repeaters could occasionally repeat, if we hit the short period tail of the FRB period distribution; and 3) series of sub-Jansky level short radio bursts could be observed as electromagnetic counterparts of the mergers of binary neutron star systems.
    }


   \maketitle

\section{Introduction} 

The origin of fast radio bursts (FRBs), these brief, coherent, and numerous radio pulses, has not been identified yet. Today, radio-astronomy surveys from all over the world have detected more than 700 FRBs, among which 137 have been officially reported \citep{Petroff2016}.

The large inferred dispersion measures (DM) 
point towards these being mostly at cosmological distances. The extragalactic origin is further confirmed by the isotropic distribution of FRBs over the sky. So far, FRBs have been detected with fluences ranging from sub-Jansky up to more than $400$\,Jy, with steep energy spectra~\citep{James_2019}. Consequently, the isotropic energy equivalent of an FRB is more than ten billion times higher than  galactic pulsar emissions, with, in addition, spectra that are radically different from most of the known radio sources. 

A fraction of FRBs appear to repeat, that is with multiple bursts spaced over  a few seconds to months,   observed at the same location. This implies that FRBs could belong to two distinct populations: repeaters and non-repeaters. Among the hundred of events published so far, about $22$ appear to repeat, mostly with no apparent periodicity, even though one has been reported to be periodic~\citep{collaboration2020periodic}. A large fraction of the repeating FRBs have been discovered by CHIME, operating at around $400$\,MHz \citep{Chime2019a,Chime2019b,Scholz2019,2019Natur.566..235C,collaboration2019chimefrb,Fonseca_2020}. The absence of real differences in their spectra, however, suggests that the two populations may originate from the same sources. 

The event rate, extrapolated from current observations that are necessarily limited in the observation time and field of view, suggests that FRBs occurs at an extraordinarily high rate of thousands per day, implying that the objects at the origin of these emissions must be numerous in the Universe~\citep{Petroff_2019}.

From a theoretical perspective, no consensual emission mechanism has been found, nor is there an accepted explanation for the two observed populations of repeaters and non-repeaters. A vast number of emission models exist, from exotic alien signals to cosmic strings, and they can be found in~\cite{Platts2018}. The recent detection of two intense radio bursts, coincident with X-ray bursts and localized at the position of SGR1935+2154, points towards the magnetar hypothesis as a source of FRBs~\citep{mereghetti2020integral,2020Natur.587...59B,collaboration2020bright}. This might, however, apply to a subset of the population only, since the equivalent luminosity of the radio bursts from SGR1935+2154 seems to be $40$ times dimmer than the dimmest FRB.

Although the number of FRB detections is growing fast, the observational constraints remain limited. The key observables at this stage, besides the energy budget and time variability, are the rates of bursts and of repeating events. These numbers are challenging to reconcile with the existing source models in the literature.\\ 

In this paper, we propose a global scenario which could explain the rates of both repeating and non-repeating events with a population of neutron stars in binary systems. 
Several studies have shown that the infall of bodies onto a compact object should lead to observable electromagnetic signals. In particular, via the Alfv\'en wing emission mechanism presented in \cite{Mottez2014}, this emission could be the source of FRBs. Other authors have proposed that FRBs result from  the impact of asteroids and comets on central compact objects \citep{Geng15, Dai_2016, Smallwood_2019}. Interestingly, the above models could naturally lead to repeating signals, as long as small bodies, such as asteroids, pass by the star at a rate corresponding to the observations. Furthermore, it provides a natural explanation to the dichotomy between repeater and non-repeater FRBs.

Such scenarios require however both a large number of progenitors, and an efficient infall mechanism into the neutron-star Roche lobe. The {\it Kozai-Lidov} gravitational effects applied to the numerous binary neutron star systems naturally provide such a framework.

In the following, we study the effect of the Kozai-Lidov mechanism on a triple system consisting of a central neutron star, a binary companion, and sizeable bodies orbiting nearby, such as an asteroid belt around the neutron star. Bodies perturbed by gravitational effects leave their orbits and fall onto the central object~\citep{Naoz_2016}. For instance in the Solar System, the Kozai-Lidov mechanism is responsible for the Kirkwood gap in the asteroid belt, under the influence of Jupiter~\citep{1943JRASC..37..187D}.  Furthermore many astrophysical systems have been found to be consistent with the implication of Kozai-Lidov perturbations: such as the formation of hot Jupiters systems via the planet-planet interactions~\citep{Naoz_2011}, the formation of close compact binaries via mass loss channels induced by secular effects~\citep{Shappee_2013,Michaely_2014}, and the pollution of white dwarf atmospheres due to the infall of asteroid and comet-like materials~\citep{Stephan_2017}.

The first discovery of earth-mass exoplanets was indeed around a millisecond pulsar~\citep{1992Natur.355..145W}. The existence of asteroid belts around millisecond pulsars has been invoked to explain various timing variations and other observational features~\citep{Cordes08,
2013ApJ...766....5S, 2014ApJ...780L..31B, 2016RAA....16...75Y,Mottez13_a}.

This study is strongly related to the one presented in~\cite{mottez2020repeating}, where the authors discuss the possible FRB emission from the interaction between an asteroid belt and a pulsar. This is why we often refer to their work regarding the radio emission mechanism. 
However, our work focuses on the orbital dynamic of the asteroids inside the belt. In this perspective, we first present the FRB emission model and the parameter sets required for the signal to be observed. We then  
compute the Kozai-Lidov time-scales for our binary system (Section~\ref{section:KL_time}) and discuss the implications in terms of FRB rates, taking into account the binary population rates (Section~\ref{section:FRBrates}). We simulate the Kozai-Lidov effect on a mock solar-like asteroid belt in Section~\ref{section:application_belt}. Finally, we discuss the broader applications of this calculation in Section~\ref{section:discussion}.

\section{FRB emission from asteroids orbiting a pulsar}\label{section:param}
Asteroid belts close to neutron stars have been previously proposed to explain observational timing and radio features \citep{Cordes08,2013ApJ...766....5S, 2014ApJ...780L..31B, 2016RAA....16...75Y,Mottez13_a,Mottez13}. No asteroid belt has yet been observed at distances larger than $1$\,A.U., but this is likely due to observational  bias. Asteroid belts could be the remains of planetary objects destroyed by the supernova that led to the formation of the neutron star, or result from the supernova fallback itself \citep{Menou2001,2008sptz.prop50345S}. The aggregation of the debris to form a planet depends mostly on external conditions \citep{Morbidelli2016}. In particular, the presence of Jupiter prevents the formation of planets in the Solar asteroid belt. The perturbations produced by an outer black hole at $\gtrsim {\rm few}$\,A.U. with a mass of $10$\,M$_{\odot}$ would be several orders of magnitude more intense than the influence of Jupiter on the Solar system belt. Therefore it is likely that no planet would form inside this asteroid belt.

\cite{Mottez2014} presented the extension of the Alfv\'en wing theory (see e.g.,~\citealp{2004jpsm.book..537S}) to relativistic winds induced by a pulsar and interacting with a companion body (e.g., planet, comet, asteroid, etc.). The emission mechanism can be summarized in three steps: first the relativistic and magnetized wind enters in direct contact with the orbiting body, creating a magnetic coupling. This direct contact induces a current sheet called an Alfv\'en wing, extending from the body far into space. Finally, the interaction of the outflow plasma crossing the Alfv\'en wing results in radio emission through coherent mechanisms such as the cyclotron maser instability.

For an asteroid of radius $R_{\rm ast}$ orbiting at distance $a_{\rm ast}$ from a pulsar located at distance $D$ from the observer, the average flux density of radio waves inside the cone of emission of opening angle $1/\gamma$, with $\gamma$ the Lorentz factor of the wind, reads~\citep{Mottez2014,mottez:hal-02569291,mottez2020repeating}:
\begin{align} \label{eq:FRB_flux}
{\langle S \rangle} =&4.3\,{\rm Jy}\,\frac{\epsilon_{\rm w}}{10^{-2}}\,{A_{\rm cone}} \nonumber\\
&\times \left(\frac{\gamma}{3\times10^6}\right)^2\left(\frac{R_{\rm ast}}{100\,{\rm km}}\right)^2 \left(\frac{a_{\rm ast}}{10^{-2}\,{\rm A.U.}}\right)^{-2} \nonumber\\
&\times \left(\frac{R_{\star}}{10^6\,{\rm cm}}\right)^{6} \left(\frac{B_{\star}}{10^{13}\,{\rm G}}\right)^{6}  \left(\frac{P_{\star}}{0.1\,{\rm s}}\right)^{-4} \nonumber\\
&\times \left(\frac{D}{100\,{\rm Mpc}}\right)^{-2}\left(\frac{\Delta f}{1\,{\rm GHz}}\right)^{-1} \ ,
\end{align}
here $\Delta f$ is the spectral bandwidth of the emission, $\epsilon_{\rm w}$ the wind power conversion efficiency, and $R_\star$, $P_\star$, $B_\star$ the pulsar radius, rotation period and dipole magnetic field strength. $A_{\rm cone}=4\pi/\Omega_A\ge 1$ is an anisotropy factor, with $\Omega_A$ the solid angle in which the radio-waves are emitted in the source frame. For an isotropic emission, $A_{\rm cone}=1$ and if, the instability triggering the radio emissions is the cyclotron maser instability, $A_{\rm cone}\sim100$ \citep{mottez2020repeating}. 

One should note that, in~\cite{mottez:hal-02569291} a revised version of the Alfv\'en wing mechanism is presented, where the magnetic flux $\Psi$ of the wind is evaluated where the field lines are wind-like and not estimated at the surface of the neutron star as previously done in~\cite{Mottez2014}. Although the physics of the process remains identical to the previous version of the study, the intensity of the radio emission is scaled down. In the present study, we use the revised version of the mechanism.

It is interesting to note that in this radio emission mechanism model, magnetar-like objects with a strong magnetic field could power FRB emission of hundreds of Janskies as observed in the ASKAP survey. Such phenomena are also suggested by the recently observed double radio bursts from the magnetar SGR1935+2154~\citep{2020Natur.587...59B,collaboration2020bright}, also coincident with X-ray bursts~\citep{mereghetti2020integral}.

In light of this emission equation, we discuss below the parameters required for the pulsar and the asteroids in order to produce an observable FRB. 

\subsection{Pulsar parameters}\label{section:pulsar_param}

Neutron stars are frequently formed in binary star systems, but the subsequent evolution of these systems leads to diverse final configurations, depending on the pre-supernova mass, the asymmetry of the explosion, a possible impulsive “kick” velocity impinged on the neutron star at birth, etc.: a parameter-space explored with sophisticated numerical simulations (e.g., \citealp{Lorimer2008, Toonen_2014} and references therein). 

We focus here on neutron star-white dwarf (NSWD), neutron star-main sequence star (NSMS), neutron star-neutron star (DNS) and neutron star-black hole (NSBH) binaries, which are found to be common outcomes of the evolution of binary systems containing neutron stars \citep{Portegies96,Nelemans2001}.

In a majority of NSBH systems, the neutron star is born with normal pulsar characteristics (e.g.,  non-recycled pulsars with large magnetic fields and mild spin periods). Various evolutionary studies show indeed that it is difficult to form recycled pulsars in these systems and their low inferred rates are compatible with their non-detection in radio so far \citep{Sipior_2004,Pfahl_2005,Shao_2018,Kruckow2018} .

The case  that, in the majority of NSWD systems, the white dwarf is formed first has also been studied numerically \citep{Toonen2018} and supported observationally (e.g., \citealp{1999MNRAS.309...26P,2000ApJ...543..321K,Manchester2000})  Hence these systems contain normal (non-recycled) neutron stars in eccentric orbits. 

Finally, observations confirm the natural scenario in which double neutron star systems contain at least one normal pulsar \citep{Tauris_2017}, which serve in our framework as the central object. 

The evolutionary path of NSMS suggests that main sequence stars should be companions to normal radio pulsars, and their (scarce) observations support this scenario (\citealp{Lorimer2008} and references therein). 

Two neutron star systems containing a planet companion have been observed (e.g., \citealp{Lorimer2008} and references therein). Data and studies on these objects are scarce, hence we mostly concentrate on the binaries mentioned above in this paper. However, we also discuss the possible contribution from these planetary systems. The two planets have been detected around millisecond pulsars, but it is impossible as yet to infer population characteristics, and normal pulsars are more numerous than millisecond pulsars and statistically likely to host planets.

For all these binary systems, it appears to be justified to assume that the neutron star presents the characteristics of a normal pulsar. We note however that the systems in close orbit, with companion semi-major axis $a_{\rm c}\ll 1\,$A.U. are usually associated with recycled pulsars. 
\\

In our model, the FRB emission happens in the first $\lesssim 10^4\,$yrs of the birth of the pulsar, and for close binaries, even within the first 10\,yrs (see Section~\ref{section:FRBrates}). The relevant pulsar parameters are hence those at birth. It is commonly accepted that the dipole magnetic field strength of the pulsar experiences little decay, with an average initial value of $10^{12.65}\,$G \citep{Faucher06}. Recent simulations show that the initial spin period could be as low as $20\,$ms \citep{Johnston_2017} and typically below $P_\star<150\,$ms \citep{Gull_n_2014}. 

The numerical values of Equation~(\ref{eq:FRB_flux}) demonstrates that such fiducial normal pulsar parameters suffice to produce  observable radio emission at the Jansky level, provided that the asteroid presents specific characteristics, which we detail in the next Section. 

We notice also that recycled pulsars, that have low magnetic fields of $B\lesssim 10^9\,$G and $P\sim {\rm few}$\,ms, are not powerful enough to produce FRB emissions at the Jansky level, except for extremely large asteroids.

\subsection{Asteroid size}\label{section:asteroid_size}
The radio emission crucially depends on the radius $R_{\rm ast}$ and orbital distance $a_{\rm ast}$ of the asteroid. One can infer from Equation~(\ref{eq:FRB_flux}) that large asteroids with radius $R_{\rm ast}\gtrsim 3\,$km are favored to power observable FRBs. From simple fragmentation arguments, it can be shown that the asteroid size distribution roughly follows a power-law~\citep{MPCOD}
\begin{align}
N_{\rm ast}\sim 100\, \left({R_{\rm ast}}/{100\,{\rm km}}\right)^{-2}\ .
\end{align}
Larger, less numerous asteroids could produce intense bursts, at a lower rate. Conversely, mJy emission, detectable with current instruments, could be produced by smaller ($3-10\,$km), more numerous asteroids. 

\subsection{Asteroid belt distance}
Equation~(\ref{eq:FRB_flux}) shows that short distances from the central neutron star are required for the body to be immersed in strong magnetic fields. Although mJy emission can be produced at a distance $a_{\rm mJy}\sim 0.1\,$A.U. from the neutron star, shorter orbital distances are required to power more intense bursts. 

The shortest possible distance corresponds to the Roche limit. The Roche limit for an asteroid falling onto a neutron star is computed to be
\begin{align}\label{eq:Roche}
&d^{\rm Roche}_{\rm NS} = 2 R_{\rm ast} \left(\frac{M_{\rm NS}}{M_{\rm ast}}\right)^{1/3} \nonumber
\\ &\sim 9.2\times 10^{-3}\, {\rm A.U.}\, \left(\frac{2\,{\rm g\,cm}^{-2}}{\rho_{\rm ast}}\right)^{1/3} \left(\frac{M_{\rm NS}}{1.4\,M_\odot}\right)^{1/3} 
\end{align}
with $R_{\rm ast}$ the asteroid radius, $M_{\rm ast}$ its mass, $\rho_{\rm ast}$ its density, and $M_{\rm NS}$ the central compact object mass. 

Asteroids could penetrate deeper than the Roche lobe if the so-called plunge factor is taken into account \citep{2016PhRvD..93d3508A}, allowing for shorter $a_{\rm ast}$ to be reached at maximum eccentricities. This would enable smaller ($R_{\rm ast}\sim 3-10\,$km) -- more numerous ($N_{\rm ast}\sim 10^{4-5}$) -- asteroids to emit Jansky-level bursts. 

We note that even at these close distances, small objects like asteroids are in general not evaporated via induction heating by the winds of the central neutron star~\citep{Kotera_Mottez16}. Their size is indeed shorter than the typical wind electromagnetic wavelength, in the framework of the Mie theory. The effects of non-sphericity, as is the case for asteroids, are $\lesssim 30\%$ on light absorption coefficients \citep{Mishchenko99}.\\

The required short orbital distances imply that, unless most asteroid belts are already created in this emission zone delimited by $d_{\rm Roche}$ and $a_{\rm mJy}$, the process of~\cite{Mottez2014} and~\cite{mottez2020repeating} can only work if asteroids actually fall close enough to the central object. We propose here that this can happen via the Kozai-Lidov effect. We set our fiducial asteroid belt distance to $a_{\rm ast}=1\,$A.U. in the following. 

We note that observations of pulsars show that there might be asteroid belts at $\sim R_\odot$ \citep{Cordes08,Mottez13_a,Mottez13}: these do not need to undergo infall in order to produce FRBs, as they are already deep into the strong wind region to produce strong Alfv\'en wing emissions. The signals from such belts could present some periodicity due to the regular orbits as observed for FRB180916, which presents a $\sim 16.35$\,days periodicity~\citep{collaboration2020periodic}. 
Indeed for favorable configurations, the alignment between the asteroid periodical motion and the observer's line-of-sight could result in a periodical observation of bursts.
However, turbulence effects in these inner wind regions along the observer's line of sight may play a role in modifying such periodicities, an effect that we do not address here.
We note also that~\cite{Jones2008} shows that infrared emission limits the inner radius of an asteroid belt to a factor that is  two or three times larger than $\sim R_\odot$.

Finally, regarding the orbital modifications of the asteroids due to the supernova phase, two cases can be distinguished: close systems where most probably the asteroid belt or debris belt forms after the supernova phases, in that case the system  is already relaxed in some way. For wide systems the large distance of the companion should not affect small bodies highly bound to the central neutron star, except for secular effects.

\subsection{Reconciling the emission beaming with the observed FRB rate}\label{section:beam}
FRB emission would be observed when the radio beam of the Alfv\'en wings crosses the observer's line of sight. This probability is diminished by the narrow emission beam (of opening angle $1/\gamma\sim 10^{-6}-10^{-5}$) produced by the Alfv\'en wave mechanism of~\cite{Mottez2014}, but compensated by the large number of orbits achieved by the asteroids before reaching the Roche limit. The time-scale for the asteroid eccentricity to shift from $a_{\rm mJy}\sim 0.1$\,AU to the Roche Limit in the emission zone (due to Kozai-Lidov effects) would be  a fraction of the Kozai-Lidov timescale, which is a secular effect, hence happening on times much larger than the orbital time period of the asteroids. Therefore the number of Keplerian orbits performed in the emission zone before reaching the Roche limit is large. A rough estimate of the number of orbits achieved by the asteroid in the emission zone can be obtain by comparing the Kozai-Lidov timescale to the orbital periods of the asteroid at the beginning of the emission zone and at the end (the Roche limit). Considering Keplerian orbits, the orbital period can be derived from Kepler's third law $P_{\rm ast} = \sqrt{a_{\rm ast}^2 GM_{\rm NS}/4\pi}$, where $P_{\rm ast}$ is the orbital period of the asteroid, $a_{\rm ast}$ its semi-major axis, $G$ the universal constant of gravitation and $M_{\rm NS}$ the mass of the orbited pulsar. For an asteroid position at the beginning of the emission zone $a_{\rm ast, mJy}\sim 0.1$\,AU, and orbiting a pulsar of mass $M_{\rm NS}\sim 1.4\,M\odot$, this period is about $P_{\rm ast, mJy}\sim 6\times 10^{3}$\,days, while at the end of the emission zone (Roche limit) it is much shorter, about $P_{\rm ast, Roche}\sim 5.5$\,h. Therefore the comparison of $t_{\rm KL}$ the Kozai-Lidov timescale, given by Eq.~\eqref{eq:tKL} (see Section~\ref{section:KL_time}), considering an outer body of mass $10\,M_\odot$ with a semi-major axis of $10$\,A.U., with the orbital periods of the asteroid gives the number of orbits achieved $N\sim t_{\rm KL}/P_{\rm ast}\sim 10^5-10^{7}$. The large number of orbits can thus compensate for the strong beaming and lead to more than one emission burst per asteroid, as we assume in the rest of our discussion. Other asteroids can also enter  the emission zone, leading to repetitions of bursts. 

In addition, turbulence effects, wind fluctuations and asteroid proper motions will also randomly affect the beam position and orientation. From~\cite{mottez2020repeating} the authors, derive a conservative value of the emission source velocity, due to the wind intrinsic oscillations, of about $v_s\sim 0.01c\ll v_{\rm wind}$, equivalent to an angular velocity of about $\dot{\omega} \sim 10^{-4}$\,rad/s. Consequently, the emission beam wanders over an area proportional to the time of observation $t_{\rm obs}$ and the Keplerian orbital period of the asteroid $n_{\rm ast} = \sqrt{GM_{\rm NS}/a_{\rm ast}^3}\sim 1.6 \,{\rm \, days}\qty(M_{\rm NS}/1.4 \, M_{\odot})\qty(a_{\rm ast}/10^{-2} {\rm \, U.A.})$, assuming the orbital motion is in the same plane as the observer line of sight for simplification. This area can be described with an opening angle $\alpha_{\rm w} = n_{\rm ast} t_{\rm obs} \sim 10^{-1} {\rm \, rad\ }\qty(M_{\rm NS}/1.4 M_{\odot})\qty(a_{\rm ast}/10^{-2}\, {\rm U.A.}) \qty(t_{\rm obs} / 1 {\rm \, h})\gg \alpha_{\rm beam} \sim \gamma^{-1}$ and defines the probable detection region. During the observation time $t_{\rm obs}$, multiple bursts can be observed if the beam crosses the observer's line of sight several times. Another consequence of the beam wandering motion is the burst duration, which result from the sweep time of the beam across the observer's line of sight, given by
\begin{align}
\tau_{\rm burst} =& \frac{\alpha_{\rm beam}}{n_{\rm ast} + \dot{\omega}} \nonumber \ ,
\\ \sim& 7{\rm \, ms} \ \qty(\frac{\gamma}{3\times10^6}) \qty(\frac{M_{\rm NS}}{1.4 M_{\odot}})\ \qty(\frac{a_{\ast}}{10^{-2} {\rm U.A.}})^{3/2}\ .
\end{align}
Finally, the number of bursts observed and their durations depend on the position of the asteroids when the emission is produced, but also on the pulsar characteristics, which make possible configurations as diverse as the observed FRB burst durations and repetitions.

Our  final picture corresponds to an emission zone filled with asteroids whose Alfven wings randomly cross the observer's line of sight during the large number of orbits achieved to reach the Roche limit, where the asteroid disruption occurs. During the disruption, complex tidal-induced fragmentation could happen, especially for large asteroids, leading to a multitude of sub-emission components over  short time-scales. Such events could explain the observations of FRB 121102, from which $\sim90$ bursts were detected during a five hour period (half falling within $30$ minutes, \citealp{Zhang2018}).

\section{Kozai-Lidov mechanisms}\label{section:KL_time}
\begin{figure*}[t!]
	\center
	\includegraphics[width = 0.8\linewidth]{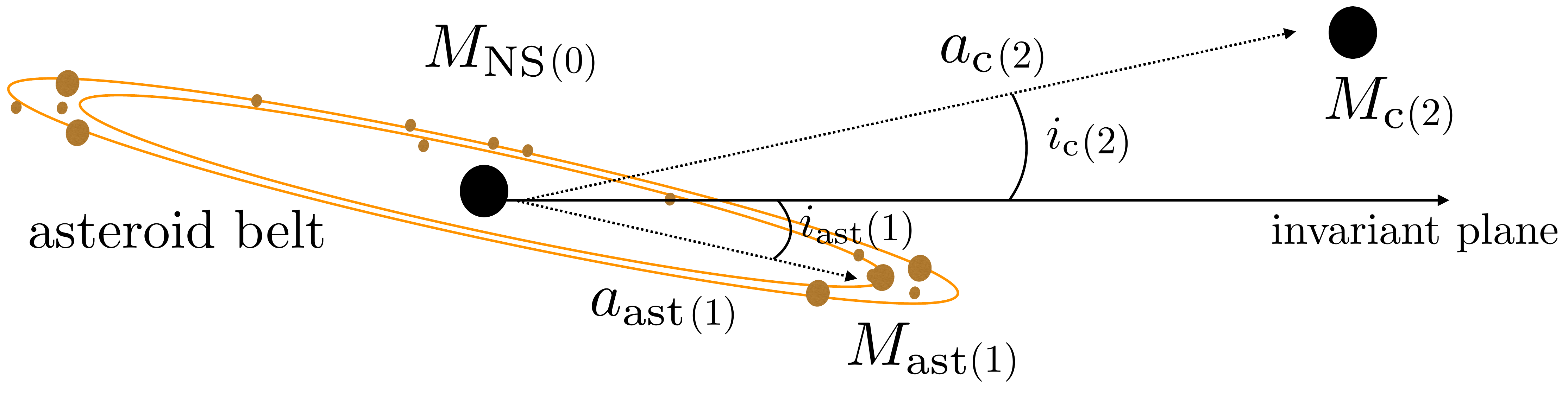}
	\caption{Framework for the Kozai-Lidov perturbation calculations in our binary neutron star+ asteroid triple system. The neutron star is surrounded by the asteroid belt, and the binary companion orbits at a larger distance. All objects are represented by their distance to the neutron star (for instance $a_{\rm ast}$ and $a_{\rm c}$) and their inclination (for instance $i_{\rm ast}$ and $i_{\rm c}$) with respect to the invariant plane.}
	\label{fig:sketch}
\end{figure*}

In the framework of asteroids orbiting a central pulsar and surrounded by an outer massive body (see Figure~\ref{fig:sketch} for a sketch), we expect modifications of the orbits to occur through the exchanges of orbital momentum between the two two-body systems: (1) pulsar-asteroid (the inner binary), and (2) (pulsar-asteroid)-outer body (the outer binary). 
These exchanges can translate into an increase of the eccentricity of the inner binary and therefore lead to configurations where the two bodies of the inner binary move very close to one another, when reaching the periapsis of their orbits, leading in some cases to a crossing of the Roche limit.

As can be seen in Figure~\ref{fig:sketch}, the subscripts $1,2,3$ refer to the central object, the outer body and the "massless" body orbiting the central object respectively. We specify the notations in some numerical estimates and in the next sections by denoting these objects with the subscripts ${\rm NS},{\rm c},{\rm ast}$, corresponding to the central neutron star, its binary companion and the orbiting asteroids.

\subsection{Secular perturbations in three-body systems} \label{section:secular_dynamics}
The motion of the outer body, also referred to as the perturbing body, induces gravitational perturbations which happen on secular timescales, that is on timescales much longer than the typical orbital timescales.
In the specific case of a hierarchical three body system, where the semi-major axis of the inner binary is much smaller than the semi-major axis of the outer binary $a_1/a_2 \ll 1$, this system is stable. Furthermore in the test particle approximation, where one of the bodies is considered "mass-less" ($m \to 0$), only the motion of this "mass-less" body is affected by the secular dynamics.
For large mass ratios within the inner binary, the inner binary orbit can flip from a pro-grade motion to a retro-grade motion by rolling over its semi-major axis. During one of these flips, the orbit passes trough an inclination of $90\degree$ which leads to a large eccentricity excitation.

The three-body dynamics is usually decomposed into the dynamics of two two-body systems, plus a perturbation effect between these two-body systems. In terms of Hamiltonian, one can write
\begin{align}
\mathcal{H} = G\frac{m_0 m_1}{2 a_1} + G \frac{\qty(m_0 + m_1)m_2}{2 a_2} + \mathcal{H_{\rm pert}} \ ,
\end{align}
where $\mathcal{H}$ is the total Hamiltonian of the three-body system, $G$ is the gravitational constant, $m$ refers to the mass, $a$ to the semi-major axis, and the subscripts $1,2,3$ to one the body or one of the two two-body systems ($1$ or $2$). Finally, $\mathcal{H_{\rm pert}}$ represent the perturbation term between the two two-body systems and can be decomposed over Legendre polynomials
\begin{align}
\mathcal{H_{\rm pert}} = \frac{G}{a_2} \sum_{j=2}^{\infty} \qty(\frac{a_1}{a_2})^j \mathcal{M}_j \qty(\frac{r_1}{a_1})^j \qty(\frac{a_2}{r_2})^{j+1} P_j \qty[\cos{\qty(\Phi)}] \ .    
\end{align}
With $r_1$ and $r_2$ the distances between the two bodies of the inner binary and outer binary respectively, $P_j$ Legendre polynomials, $\Phi$ angle between $\vec{r}_1$ and $\vec{r}_2$, and $\mathcal{M}_j = m_0 m_1 m_2 [{m_0^{j-1} - \qty(-m_1)^{j-1}}]/{\qty(m_0 + m_1)^j}$ a mass term.

It is possible to rewrite this series only for the two main terms
\begin{align}\label{eq:Hpert}
\mathcal{H_{\rm pert}} \approx \mathcal{H_{\rm quad}} + \epsilon \mathcal{H_{\rm oct}} \ ,
\end{align}
where $\mathcal{H_{\rm quad}}$ and $\mathcal{H_{\rm oct}}$ represent the quadrupolar and octupolar orders of the perturbation and $\epsilon$ is given by
\begin{align} \label{eq:epsilon}
\epsilon = \frac{a_1}{a_2} \frac{e_2}{1-e_2^2} \ .
\end{align}
Depending on the configuration of the three-body system, the value of $\epsilon$ indicates which order dominates the dynamic (either quadrupolar or octupolar). Furthermore stable systems are expected for values of epsilon $\epsilon \sim 0.1$ or if the eccentricity is null $a_1/a_2 \sim 0.1$.
\begin{figure*}[tb!]
	\center
	\includegraphics[width = 0.99\columnwidth]{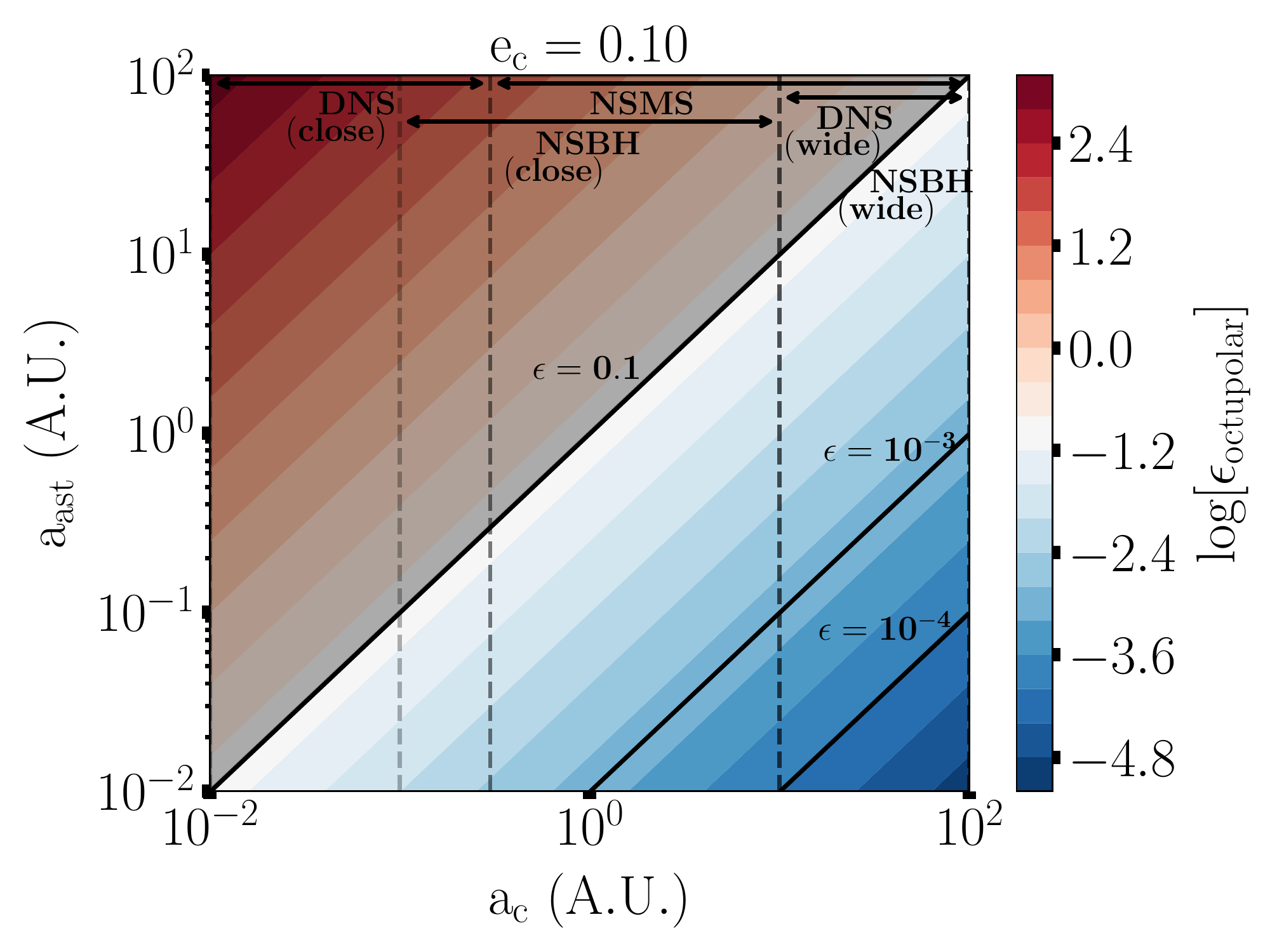}
	\includegraphics[width = 0.99\columnwidth]{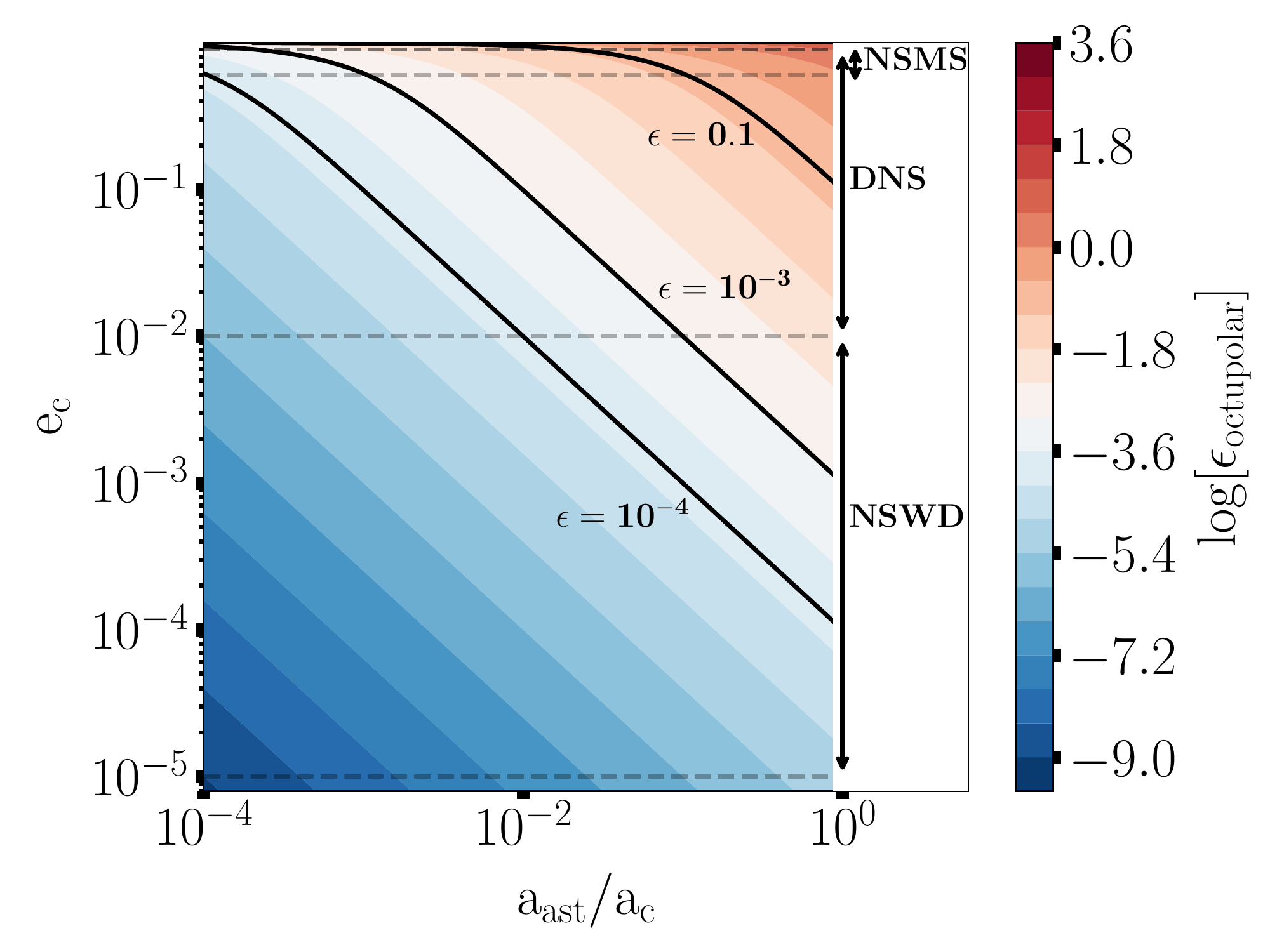}
	\caption{Evolution of the octupolar efficiency parameter $\epsilon$ (Equation~\ref{eq:epsilon}) as a function of the asteroid belt semi-major axis $a_{\rm ast}$ and the companion semi-major axis $a_{\rm c}$ ({\it left}), and the companion eccentricity $e_{\rm c}$ and ratio of asteroid belt semi-major axis $a_{\rm ast}$ and companion semi-major axis $a_{\rm c}$ ({\it right}). The solid black lines delimit the regions where the octupolar regime of the three-body dynamic is expected to be dominant $\epsilon=0.1$ and $\epsilon=10^{-3}$, and where a transition from octupolar regime to quadrupolar regime is expected to take place $\epsilon=10^{-4}$.}
	\label{fig:epsilon}
\end{figure*}
Figure~\ref{fig:epsilon} shows the evolution of the $\epsilon$ parameter depending on the inner binary (pulsar-asteroid in our case) configuration and the outer binary (perturbing body) configuration. The shadowed region represent the configurations where the outer body is closer than the inner binary (between the pulsar and the asteroid in our case), which is not possible. The domain where the octupolar is fully dominant is delimited by the two solid black lines. One can see that this region corresponds to configurations where the outer body has an eccentric orbit and is not too far from the inner binary.

When the outer body has a circular orbit, the dynamics is led by the quadrupolar term and results in the so-called classical Kozai-Lidov mechanism. In this mechanism, periodical exchanges of orbital momentum between the two two-body systems lead to a reduction of the inclination of the inner binary at a cost of an increase in eccentricity. These oscillations stem from the fact that in the test particle approximation (where one of the bodies of the inner binary has a mass close to zero), the $z$-component of the total orbital momentum, defined by the invariant plane, is conserved and can be rewritten as a function of the inclination and the eccentricity
\begin{align} \label{LZ}
L_{z} = {\rm Constant} = \sqrt{1 - e_1^2} \cos{i_{\rm tot}} \ ,
\end{align}
where $L_{z}$ is the $z$-component of the total orbital momentum of the three-body system, $e_2$ is the eccentricity of the inner binary and $i_{\rm tot}$ is the total inclination of the system within the invariant plane. By conservation principles, it is straightforward to extract the maximal eccentricity reachable as a function of the initial inclination, assuming a total transfer of the inclination during the Kozai-Lidov effect. Therefore one can obtain
\begin{align}
e_{\rm 1, max, KL} = \sqrt{1 - (5/3) \cos^2{i_{\rm tot}}}\ .
\end{align}
From the same argument, it is possible to derive the minimal initial inclination required to trigger classical Kozai-Lidov effects, which is $40\degree < i_{\rm init, tot} < 140\degree$.

In the case where the outer body has a non zero eccentricity ($e_2 \neq 0$), the octupolar term of the perturbation become non-negligible. In particular in the regime where $ 10^{-3} \lesssim \epsilon \lesssim 0.1$, the octupolar term is dominant. In this regime, the Kozai-Lidov effects are called the eccentric Kozai-Lidov mechanism (EKM), where the classical Kozai-Lidov oscillations of the quadrupolar regime are modulated by a rotation of the orbits of the inner binary around its semi-major axis. This rotation leads to an increase of the inclination of upto $i_{1}=90\degree$ beyond which the orbit flips from a pro-grade motion to a retro-grade motion. During the whole process, classical Kozai-Lidov oscillations continuously occur with intensity peaking when the orbit reaches a $90\degree$ inclination, triggering extreme eccentricities $e_{\rm 1, max, EKL} \to 1$. 
The criteria for orbits to flip has been derived by~\cite{Li_2014}
\begin{align}
    \epsilon > \frac{8}{5} \frac{1 - e_1^2}{7 - e_1 \qty(4 + 3 e_1^2)\cos{\qty(\Omega_1 + \omega_1)}} \ ,
\end{align}
where $\Omega_1$ and $\omega_1$ are the longitude of the ascending node and the argument of the periapsis of the inner binary respectively. Numerical results are consistent with this criteria (\cite{Li_2014}), and show once again how the $\epsilon$ parameter can be used to discriminate between the different dynamical regimes of the three-body system.

The EKM is characterized by longer timescales than the classical Kozai-Lidov effect, since it can be seen as the superposition of several classical Kozai-Lidov oscillations, but it leads to extremely high eccentricities. The intensity of the oscillations in the EKM depends on the value of $\epsilon$ and so on the dynamical regime of the three-body system. The EKM has been found to be possible for at least two distinct regimes: (i) Low eccentricity-High inclination, and (ii) High eccentricity-Low inclination. The first regime corresponds to the classical criteria on the initial inclination to trigger Kozai-Lidov oscillations ($40\degree < i_{\rm init, tot} < 140\degree$), and more interestingly, the second regime corresponds to orbital configurations where the system can be almost coplanar but still trigger EKM thanks to the high eccentricity of the inner binary.

In the specific framework of Kozai-Lidov effects, the three-body dynamics can be described with three main regimes: the quadrupolar regime when the outer body has a circular trajectory, featured by classical Kozai-Lidov oscillations; the octupolar regime when $ 10^{-3} \lesssim \epsilon \lesssim 0.1 $, enabling a richer dynamics with Eccentric Kozai-Lidov mechanisms and orbital flips; and finally, a combination of the previous two regimes where $\epsilon \lesssim 10^{-3}$, which depends on the specific configuration of the three-body system and is difficult to analyze in a general framework. In this study we consider the octupolar regime down to $\epsilon=10^{-4}$, where in fact a transition towards the quadrupolar regime operates. This choice is made for illustrative purposes, in order to map a larger parameter space (matching astrophysical objects) without falling into too much purely dynamical considerations. However in Appendix~\ref{app:quadrupolar}, we provide a study focused on the quadrupolar regime, showing that the conclusions drawn from the octupolar regime also hold in this regime.

\subsection{Kozai-Lidov timescales}
As described before, the dynamics of each regime, quadrupolar or octupolar, is different and so are their characteristic timescales.

Interestingly in the quadrupolar regime (and the test particle approximation), the dynamics is fully integrable, meaning that the Hamiltonian equations of motion can be solved. In this perspective, \cite{Antognini_2015} derive the exact classical Kozai-Lidov period and study its behavior across the parameter space of the three-body dynamics. In particular, it is shown that this exact period only varies within a factor of a few from the standard (and well-known) Kozai-Lidov timescale formula. It is worth noting that this is only true in general conditions, away from the boundary between the libration and rotation regime, where non-secular effects are expected, as well as away from orbital resonances.
This timescale is given by
\begin{align} \label{eq:tKL}
t_{\rm KL} \sim \frac{16}{15} \frac{a_{\rm 2}^3}{a^{3/2}_{\rm 1}} \qty(1 - e_{\rm 2}^2)^{3/2} \frac{1}{\sqrt{G}}\frac{\sqrt{m_0 + m_1}}{m_2} \ .
\end{align}

In the octupolar regime, however, the dynamics is no longer integrable, as previous quantities are no longer integrals of motion, therefore the Hamiltonian equations of motion cannot any more be solved.
\cite{Antognini_2015} shows that the exact period for the EKM can also be derived and this exact period can be well approximated with a Kozai-Lidov timescale in the EKM regime.
This new time scale is given by
\begin{align} \label{eq:tEKM}
t_{\rm EKL} &\sim \frac{128 \sqrt{10}}{15 \pi \sqrt{\epsilon}} t_{\rm KL, i=90\degree}\\
&\sim  4.8\,{\rm yr}\,\epsilon^{-1/2} \left(\frac{a_{\rm ast}}{0.5\,{\rm A.U.}}\right)^{-3/2}  \left(\frac{a_{c}}{{\rm A.U.}}\right)^3 \nonumber
\\ &\times\left(\frac{M_{\rm c}}{1\,M_\odot}\right)^{-1}\left(\frac{M_{\rm NS}}{1.4\,M_\odot}\right)^{1/2}\ ,
\label{eq:tEKM_num}
\end{align}
where $t_{\rm KL, i=90\degree}$ represents the classical Kozai-Lidov timescale and we suppose that inclinations up to $90\degree$ can be reached thanks to the orbital flip mechanisms described earlier in section~\ref{section:secular_dynamics}.
Furthermore the time-scale of Equation~\eqref{eq:tEKM} describes a full EKM cycle, with two flips: from pro-grade to retro-grade and back again.

The numerical values given in Equation~(\ref{eq:tEKM_num}) correspond to a mildly close NSMS, NSWD or DNS case, with $a_{\rm c}$ the semi-major axis of the orbiting companion and $M_{\rm c}$ its mass. The  estimate assumes a null eccentricity $e_{\rm c}=0$.

\begin{figure*}[tb!]
	\center
	\includegraphics[width = 0.99\columnwidth]{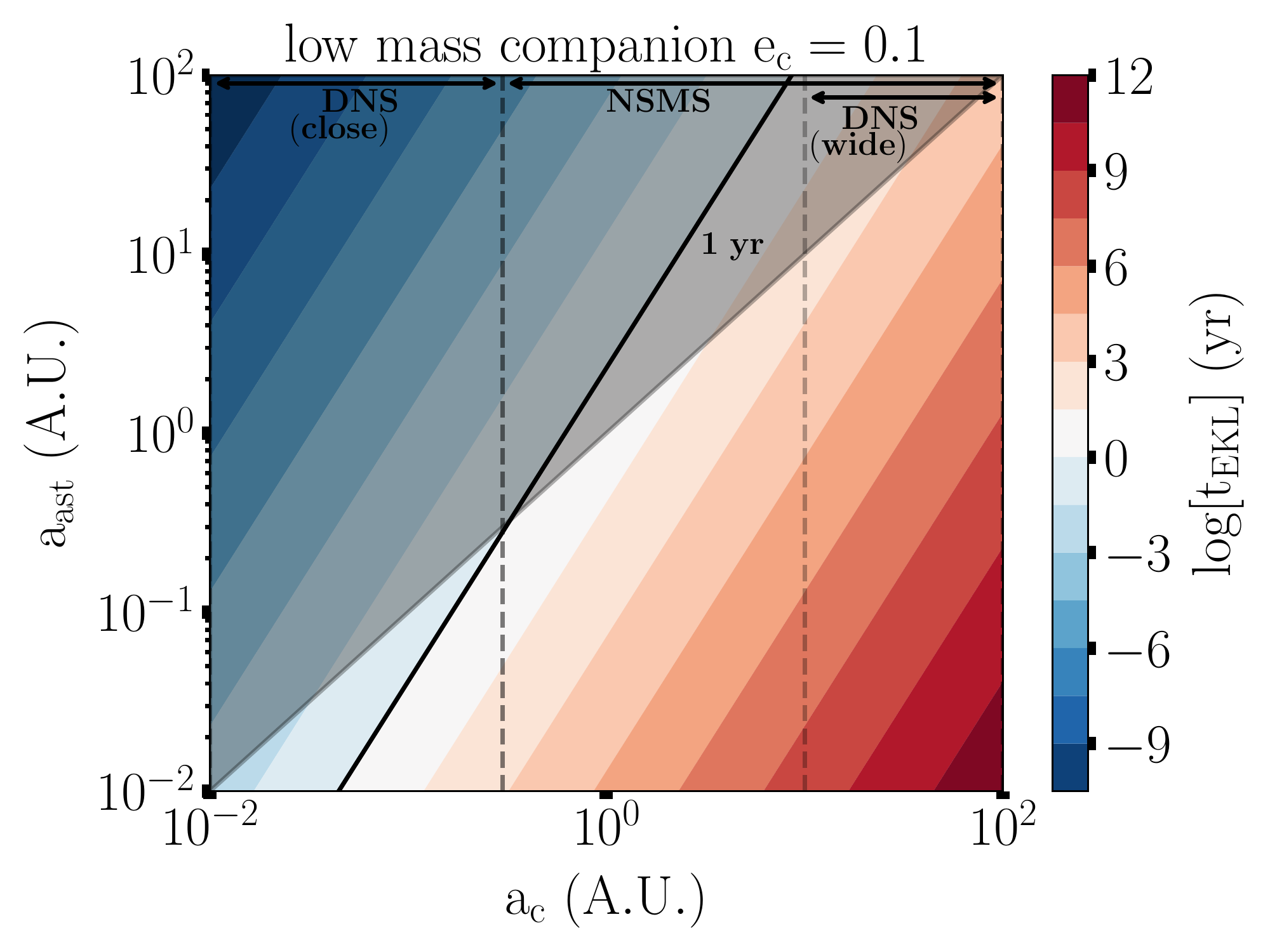}
	\includegraphics[width = 0.99\columnwidth]{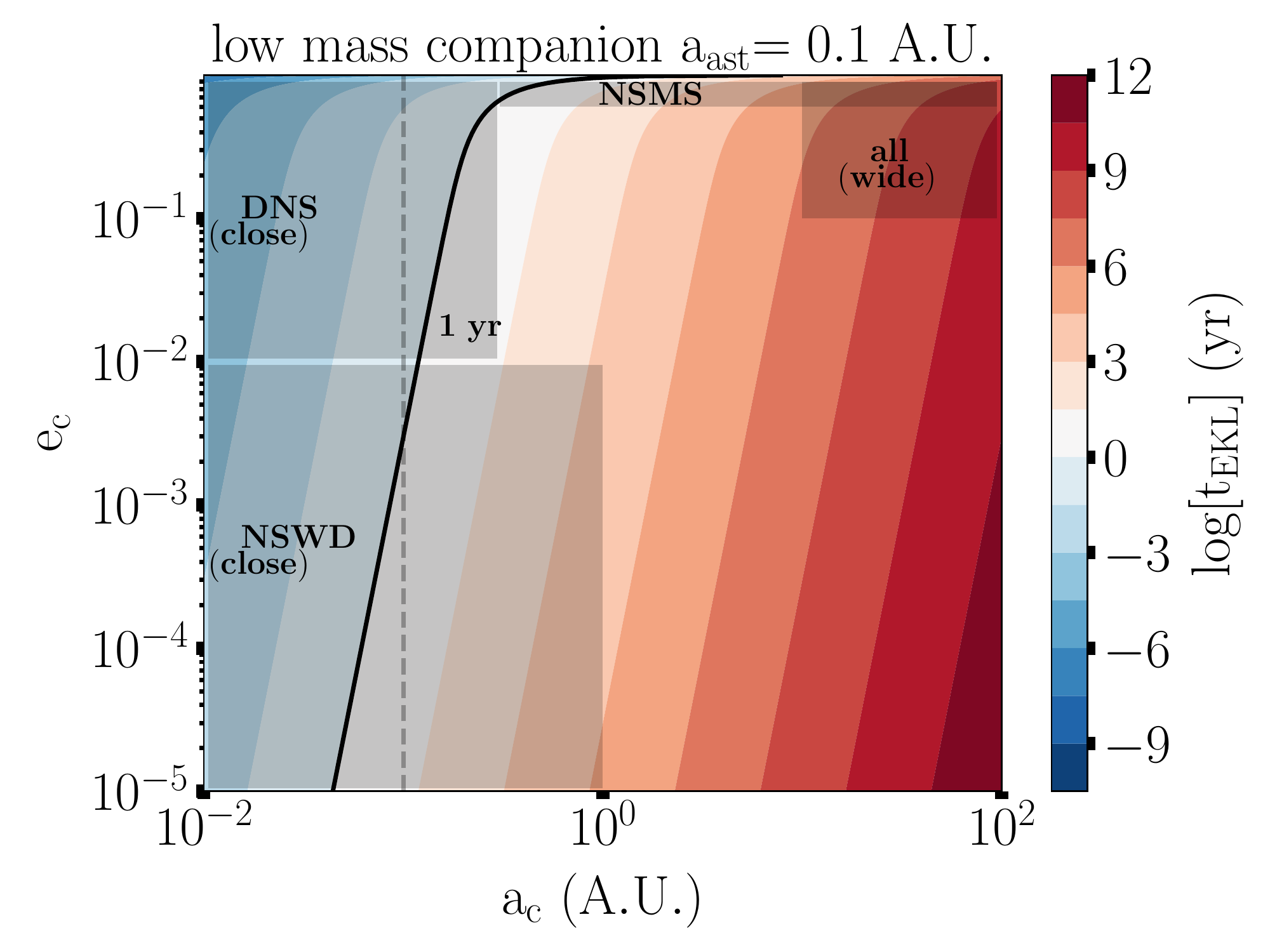}
 \\ \includegraphics[width = 0.99\columnwidth]{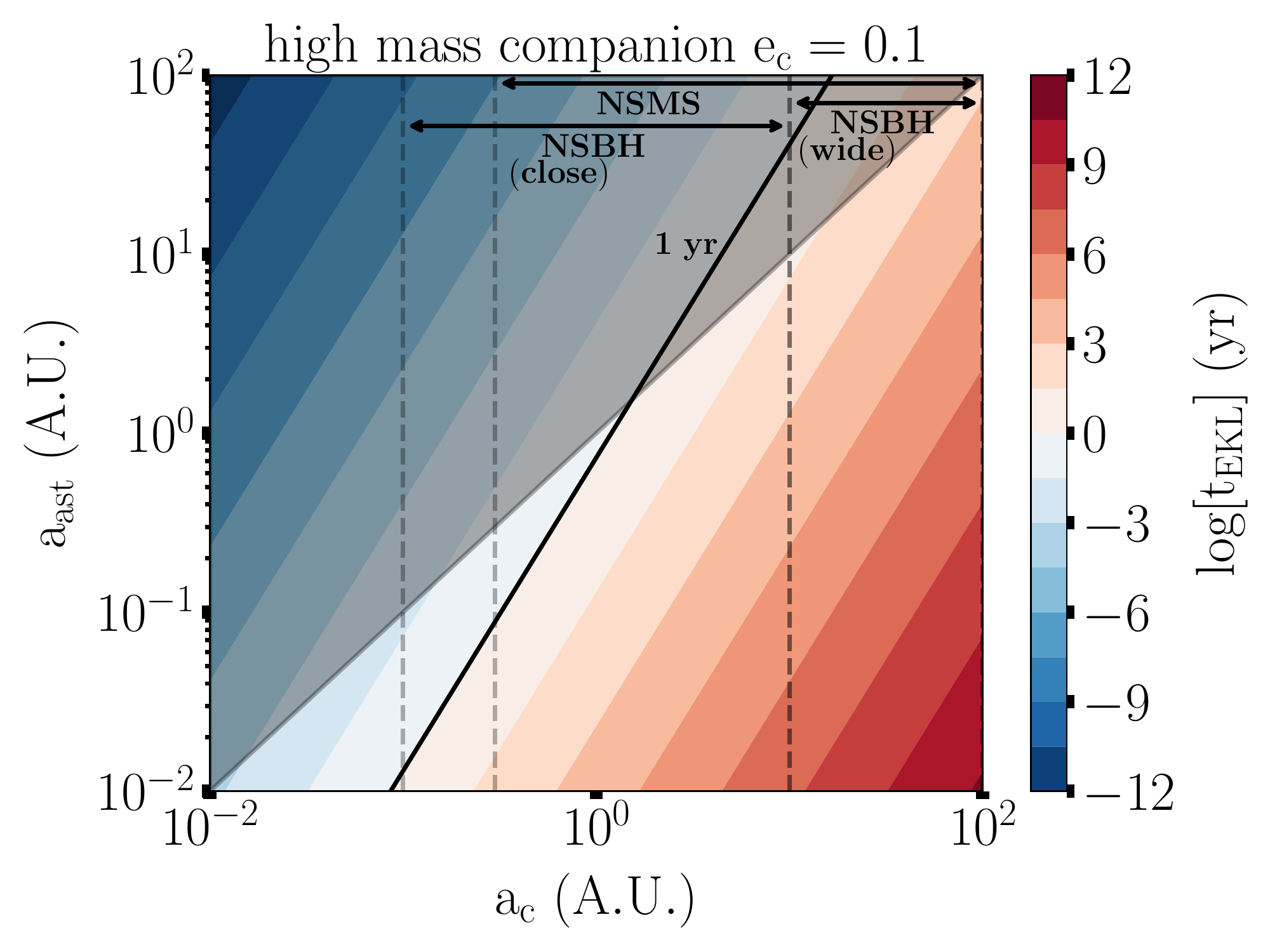}
	\includegraphics[width = 0.99\columnwidth]{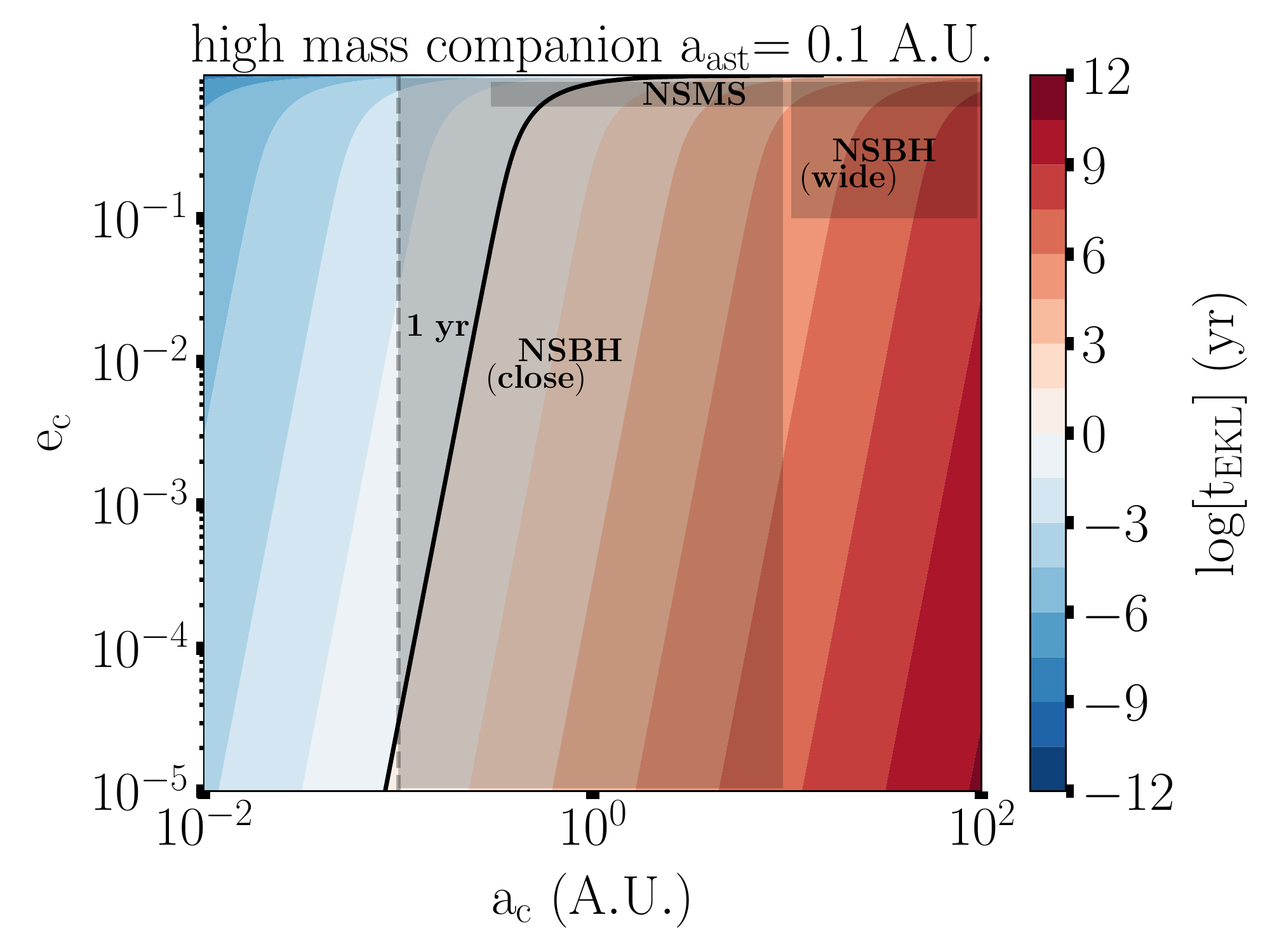}
	\caption{Evolution of the EKM timescale (given by Equation~\ref{eq:tEKM}) for various configurations of three-body systems, as a function of the semi-major axis, for a companion eccentricity $e_{\rm c}=0.1$ ({\it left}), and outer binary eccentricity and semi-major axis, for an asteroid belt located at $a_{\rm ast}=1\,$A.U. ({\it right}), for a low mass companion $M_c=1\,M_\odot$ corresponding to DNS, NSMS and NSWD systems ({\it top}), and a high mass companion $M_c=10\,M_\odot$ corresponding to NSMS and NSBH systems ({\it bottom}). The shadowed region on the {\it left} panels represents the forbidden configurations where the outer binary is closer than the inner binary. On the {\it right} panels, these forbidden configuration lies on the left-hand side of the vertical dashed line. The gray boxes on the {\it right} panels represent the parameters spaces where the different systems considered here are expected to lie. The solid lines show the limit where the timescale is shorter than one year (leading to transient events). On the {\it left} panels, the gray thin horizontal dashed lines and the arrows indicate the parameter-space in which the different systems would lie (Table~\ref{table:DNSchar}).}
	\label{fig:tEKL}
\end{figure*}

Figure~\ref{fig:tEKL} presents the EKM time-scales for various three-body system configurations. Again, the shadowed region delimits the forbidden configurations. Generically, the time-scale increases with the inner binary orbital width and with the distance of the outer binary, as expected from gravitational considerations: the farther the outer body, the lighter the gravitational perturbation on the inner binary, and similarly with the width of the inner binary.
\begin{figure*}[ht]
	\center
	\includegraphics[width = 0.99\columnwidth]{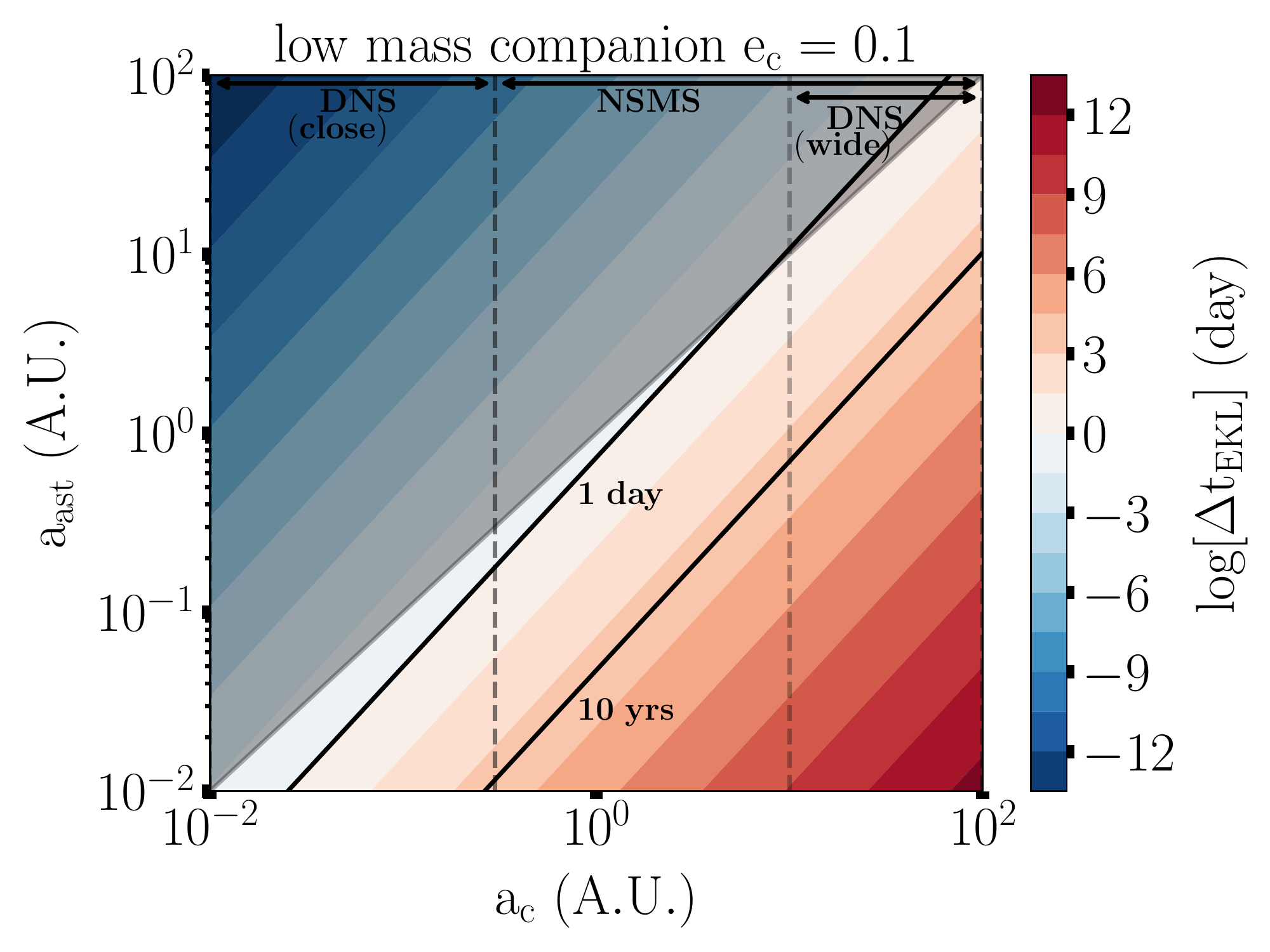}
	\includegraphics[width = 0.99\columnwidth]{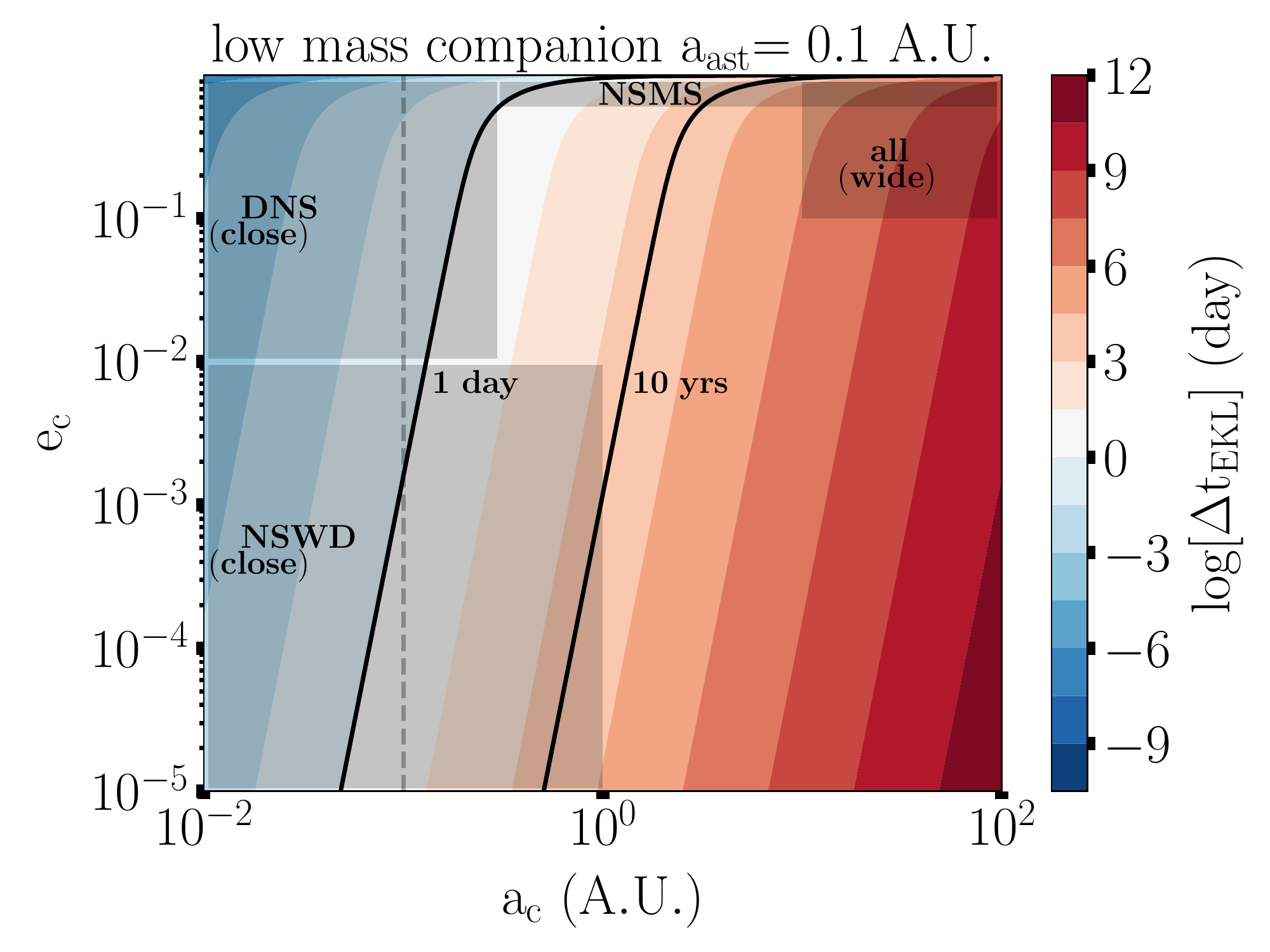}
\\ \includegraphics[width = 0.99\columnwidth]{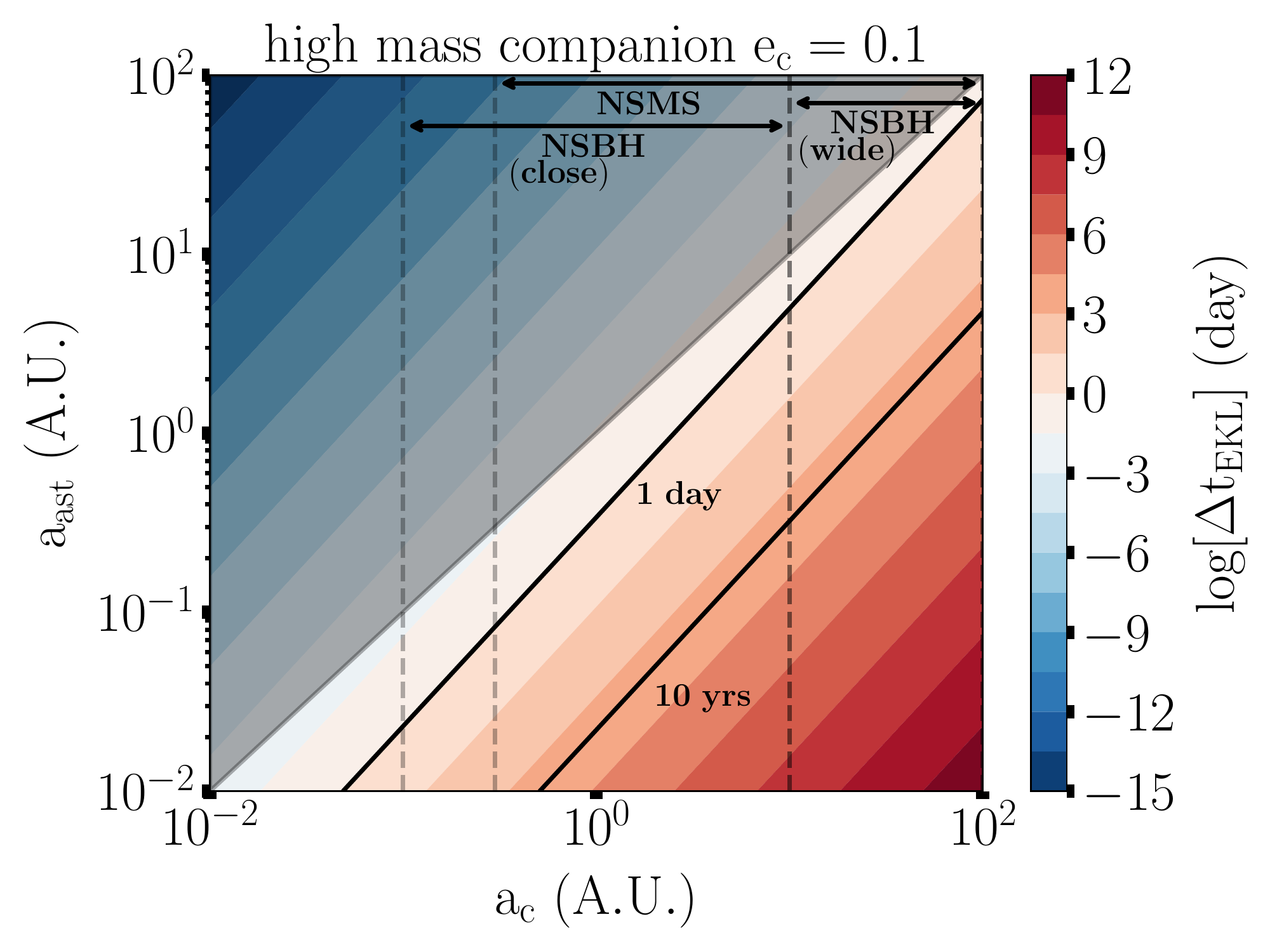}
   \includegraphics[width = 0.99\columnwidth]{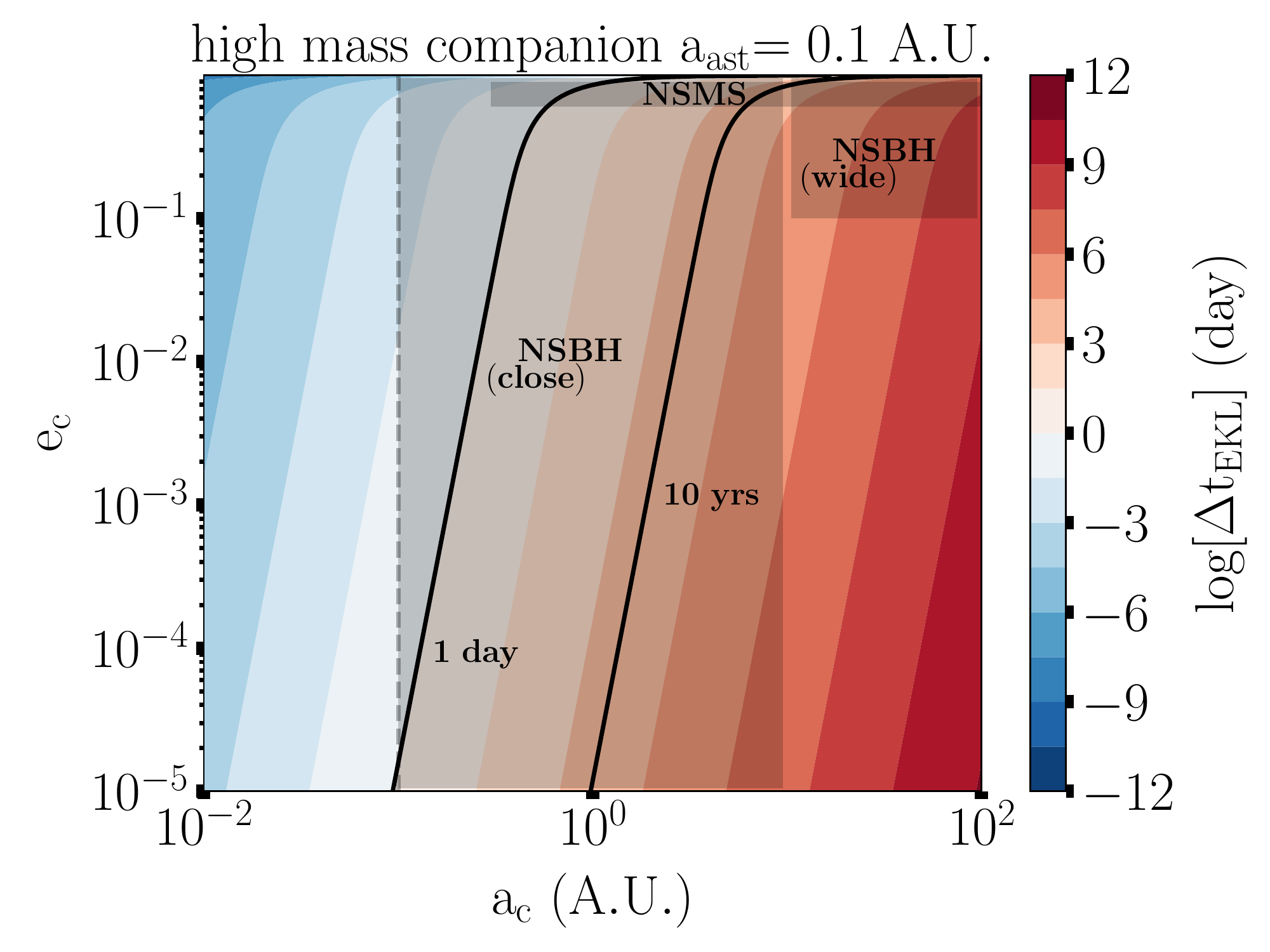}
	\caption{Same as Figure~\ref{fig:tEKL}, but for the relative Kozai-Lidov time delay  (given by Equation~\ref{eq:dtEKL}). The solid lines represent the limits where the relative delay equals $1$\,day (for sources producing day-repeaters) and the dashed line is the limit where the relative delay equals $10$\,years (observational time beyond which sources cannot be observed as repeaters).}
	\label{fig:dtEKL}
\end{figure*}

\subsection{Kozai-Lidov relative time delays}\label{section:DtKL}
In the specific case where the inner binary is in fact made of several small objects such as an asteroid belt or a comet cloud, orbiting a more massive central body, an additional interesting quantity is the relative time delay of the Kozai-Lidov time-scales between close-by objects.
The relative time delays between two small objects seperated by a distance $\Delta a_1$ is given in the quadrupolar regime by
\begin{align} \label{eq:dtKL}
\Delta t_{\rm KL} &=  \frac{8}{5} \frac{a_2^3}{a^{5/2}_1} \Delta a_1 \qty(1 - e_2^2)^{3/2}\frac{1}{\sqrt{G}} \frac{\sqrt{m_0 +m_1}}{m_2} \ , \nonumber
\\ &= \frac{3}{2}t_{\rm KL} \frac{\Delta a_1}{a_1}
\end{align}
and in the octupolar regime by
\begin{align} \label{eq:dtEKL}
    \Delta t_{\rm EKL} = \frac{256 \sqrt{10}}{15 \pi} \frac{t_{\rm KL, i=90\degree}}{\sqrt{\epsilon}} \frac{\Delta a_1}{a_1} \ .
\end{align}
Two objects separated by a distance $\Delta a_1$ orbiting a central more massive object and perturbed by an outer body, undergo Eccentric Kozai-Lidov effects with a time-scale difference given by Equation~\eqref{eq:dtKL} and~\eqref{eq:dtEKL} depending on the dynamical regime.

Equation~\eqref{eq:dtEKL} can be also rewritten in a more compact way
\begin{align}
     \Delta t_{\rm EKL} = 2 t_{\rm EKL} \frac{\Delta a_1}{a_1} \ .
\end{align}
This formula provides a more straightforward description of the relative time delay of the EKM.

Assuming that the initial distribution of $a_{\rm ast}$ in the asteroid belt follows a Normal distribution with mean $\langle a_{\rm ast}\rangle$ and width $\sigma_a=\varepsilon_{\rm ast}\langle a_{\rm ast}\rangle$, the mean distance between two consecutively falling asteroids can be estimated statistically as $\langle\Delta a_{\rm ast}\rangle\approx \sigma_a/N_{\rm ast,KL}$, with $N_{\rm ast,KL}$ the number of asteroids undergoing Kozai-Lidov effects. In the octupolar regime, when the outer body has a non circular orbit, most asteroids undergo Kozai-Lidov effects and reach high eccentricities. Hence one can write $N_{\rm ast,KL}\sim \epsilon_{\rm eff} \,N_{\rm ast}$, with $N_{\rm ast}$ the total number of asteroids in the belt and $\epsilon_{\rm eff}\gtrsim 0.2$ (see Section~\ref{section:asteroid_infall_rates}). The fraction of asteroids meeting the Kozai-Lidov criterion in the quadrupolar regime is calculated in Appendix~\ref{app:fraction_KL}. One can then express the mean relative Eccentric Kozai-Lidov time delay as 
\begin{align}\label{eq:mean_dtEKL}
\langle\Delta t_{\rm EKL} \rangle &\sim 2.7\,{\rm days}\,\epsilon^{-1/2}\epsilon_{\rm eff}\,\left[\frac{N_{\rm ast}}{100}\right]^{-1}\frac{\varepsilon_{\rm ast}}{0.15}\left(\frac{\langle a_{\rm ast}\rangle}{0.5\,{\rm A.U.}}\right)^{-3/2} \nonumber\\
&\times \left(\frac{a_{\rm c}}{{\rm A.U.}}\right)^3\left(\frac{M_{\rm c}}{10\,M_\odot}\right)^{-1}\left(\frac{M_{\rm NS}}{1.4\,M_\odot}\right)^{1/2}\ ,
\end{align}
where the numerical estimates are presented again for a mildly close NSMS or NSWD or DNS case.

Figure~\ref{fig:dtEKL} describes the evolution of the relative time delays across the parameter space allowed for the three-body system. The time delay trend follows the EKM timescales as depicted in Figure~\ref{fig:tEKL}.

\section{FRB rates for close and wide NS binaries}\label{section:FRBrates}
In this section, we apply the formalism derived in the previous Sections to populations of neutron star binaries and derive corresponding FRB rates.

\begin{table*}
\begin{center}
\resizebox{\textwidth}{!}{%
\begin{tabular}{lrrrrrr}
\toprule  \textbf{Systems} &  \begin{tabular}{@{}c@{}}$\dot{\nu}_{\rm c}$ \\$[10^{-5}\,{\rm yr}^{-1}]$\end{tabular}& \begin{tabular}{@{}c@{}}$M_{\rm c}$ \\$[M_\odot]$ \end{tabular}&\begin{tabular}{@{}c@{}}$e_{\rm c}$\\ \end{tabular} & \begin{tabular}{@{}c@{}}$a_{\rm c}$ \\$[$A.U.$]$ \end{tabular} &\begin{tabular}{@{}c@{}}$\dot{n}_{\rm FRB}$ \\$[10^3\,{\rm Gpc^{-3}\,yr^{-1}}]$ \end{tabular}& References\\

\midrule
NSWD &	 $24$	& $0.01-1$ & $10^{-5} - 10^{-2}$ & $10^{-3}-{\rm few}\,10{\rm s}$ &  \begin{tabular}{@{}r@{}} $1.3$ (rep.)\\$96$ (nrep.)  \end{tabular} &
\begin{tabular}{@{}r@{}}\cite{Nelemans2001,Kroupa2011}\\ \cite{Lorimer2008,Hobbs_2004} \end{tabular}  \\
\midrule
NSMS &	\begin{tabular}{@{}r@{}}5.8 (all)\\ $1.7$ (wide)\end{tabular}	& few $-10$ & $0.6-0.9$ & $0.3-{\rm few}\,10$s &  \begin{tabular}{@{}r@{}} $1.2$ (rep.)\\$23$ (nrep.)  \end{tabular} &\begin{tabular}{@{}r@{}}\cite{Portegies96}, \cite{Lyne2005}\\\cite{Kaspi94}, \cite{Hobbs_2004}\end{tabular}  \\
\midrule
DNS& \begin{tabular}{@{}r@{}}$5.7$ (all)\\2 (wide) \end{tabular}	& $\sim 1.4$       & $10^{-2} - 0.9$     & \begin{tabular}{@{}r@{}}$5\times10^{-3}-0.3$ (close)\\few $-$ {\rm few}\,10s (wide)\end{tabular}&  \begin{tabular}{@{}r@{}} $0.38$ (rep.)\\$27$ (nrep.)  \end{tabular} &\begin{tabular}{@{}r@{}}\cite{Nelemans2001}, \cite{Hobbs_2004} \\\cite{Portegies_Zwart_1999}\\\cite{Tauris_2017}, \cite{Kruckow2018} \end{tabular}\\
\midrule
NSBH& $0.06-1.3$	& $\sim 5 - 100$ & \begin{tabular}{@{}r@{}}any (close)\\ $\sim 1$ (wide)\end{tabular}  & \begin{tabular}{@{}r@{}}$0.1-{\rm few}$ (close)\\few $-$ {\rm few}\,10s (wide) \end{tabular}  &  \begin{tabular}{@{}r@{}} $0.2$ (rep.)\\$1.3$ (nrep.)  \end{tabular} &\begin{tabular}{@{}r@{}}\cite{Kruckow2018}\\\cite{Kroupa2008}\\ \cite{Belczynski99} \end{tabular}\\
\bottomrule
\end{tabular}
}
\caption{Population characteristics (binary system Galactic birth rate $\dot{\nu}_{\rm c}$ and companion mass $M_{\rm c}$) and orbital element distributions (eccentricity $e_{\rm c}$ and semi-major axis $a_{\rm c}$) for neutron stars in binary systems: neutron star-white dwarf (NSWD), double neutron star (DNS), and neutron star-black hole (NSBH). The FRB rate densities  $\dot{n}_{\rm FRB}$ are estimated using Eqs.~(\ref{eq:FRBrep}-\ref{eq:FRBrate}), see Section~\ref{section:wide_close}.}
\label{table:DNSchar}
\end{center}
\end{table*}

\subsection{Neutron star binary system population characteristics}\label{section:binary_char}
A neutron star is formed among a binary stellar system when the initially more massive star undergoes a supernova explosion. The companion can be a main sequence star, have already transformed into a white dwarf, or become a neutron star or a black hole following a second supernova explosion \citep{Portegies96, Lorimer2008}. However, in most scenarios, the explosion or the kick experienced by the neutron star at birth disrupts the binary system~\citep{Hansen_1997,Lu_2019}.

The majority of stellar binaries are initially wide \citep{Kroupa2008,Kroupa2011}, with orbital separation $a_{\rm c}\gtrsim {\rm few}\,$A.U., and each object evolves mostly as single stars~\citep{2014LRR....17....3P}. Supernova kicks drastically reduce the rate of these wide binaries by disrupting them. Orbits with higher eccentricity are more likely to survive these kicks.

Numerical binary population synthesis indicate that systems containing main sequence stars could be of order $\nu_{\rm NSMS}\sim 5.8\times 10^{-5}\,{\rm yr}^{-1}$ \citep{Portegies96} in the Galaxy. These authors also show that wide NSMS systems represent about $(6.7/65)\%\sim 0.1\%$ of the total neutron star population (hence $\sim 1.7\times 10^{-5}\,{\rm yr}^{-1}$). The severe population cut by a factor of 65 compared to a produced number of wide binaries is due to the supernova kicks. 

Observations concur, pointing to mildly close systems with $a_{\rm c}\sim 1-{\rm few}\,$A.U., with mild to high eccentricities $e_{\rm c}\sim 0.6-0.9$ (e.g., PSR1259-63 and its $10\,M_\odot$-mass Be-star companion, \citealp{Johnston1992}, J1740-3052 and its B-type star companion of mass $M_{\rm c} \sim 11\,M_\odot$, \citealp{Madsen2012,Hobbs_2004}, PSR J1638-4715 and its $4.5\,M_\odot$-mass companion, \citealp{Lyne2005}, PSR J0045-7319 and its $4\,M_\odot$-mass companion, \citealp{Kaspi94}). 

NSWD systems are naturally more numerous, as white dwarfs are common outcomes of main sequence stars, with simulated rates $\sim 4$ times higher than for NSMS \citep{Nelemans2001}. Due to their formation channels, NSWD are frequently found in very close circular systems, in which case the neutron star is a recycled pulsar. The orbital semi-major axis distribution of NSWD binaries should however follow the same trend as NSMS systems, with 1/3 of wide binaries with high eccentricities. 

For DNS, \cite{Nelemans2001} estimate a total Galactic population rate of $\nu_{\rm DNS, all}\sim 5.7\times 10^{-5}\,{\rm yr}^{-1}$, which includes binaries with recycled pulsars which are particularly close. \cite{Portegies_Zwart_1999} calculated numerically that wide systems with neutron stars that evolved mostly independently constitute again about a third of the total DNS population, with a rate of $\nu_{\rm DNS, wide}\sim 5.7\times 10^{-5}\,{\rm yr}^{-1}$ (see also \citealp{Kruckow2018}).

For NSBH, the birth rate is estimated to $0.6-13\,$Myr$^{-1}$ in the Galactic disk \citep{Shao_2018,2013ApJ...779...72D,Lamberts18}. Recent simulations show that about $10$\% of the binary population could be wide binaries \citep{Belczynski99, Kruckow2018}. \\

The orbits of close binary neutron star systems that have low-mass companions, such as low-mass white dwarfs ($M_{\rm c}\lesssim 0.7\,M_\odot$) tend to be circular: $e_{\rm c} \sim 10^{-5}-10^{-2}$. Close systems with high-mass companions, such as neutron stars, some white dwarfs and main sequence stars ($M_{\rm c}\lesssim 0.7\,M_\odot$) have more eccentric orbits $e_{\rm c} \sim 10^{-2}-0.9$ (e.g., \citealp{Lorimer2008,Hobbs_2004}). NSBH systems have a wide range of eccentricities, that essentially span the full physically allowed range \cite{Kruckow2018}. Wide systems have highly eccentric orbits. \\

We consider for our systems, masses of $M_{\rm NS} = 1.4\,M_\odot$ (since this is the minimal mass required to produce a NS) for the central neutron star. The companion masses span over $M_{\rm c}=0.01-10\,M_\odot$ for white dwarfs to black holes. For illustration, we use $M_{\rm c}=10\,M_\odot$, a typical value in NSBH  \citep{Kruckow2018} and NSMS systems. Estimates can easily be scaled for larger black hole masses, which would lead to higher asteroid infall rates. \\

Table~\ref{table:DNSchar} summarizes the typical parameter ranges discussed above for our binary populations. 

We mentioned in Section~\ref{section:pulsar_param} that systems with planet companions have also been observed \citep{Lorimer2008}, but the rates and characteristics of these systems are not yet  clear. A planet was detected in the triple Pulsar System PSR B1620-26, with a wide inferred semi-major axis of $a_{\rm c}\sim 23\,$A.U. and moderate orbital eccentricity \citep{Sigurdsson2003}. Three planetary bodies were found orbiting at $\sim\,$A.U. distances around pulsar B1257+12 \cite{Wolszczan92}. In both systems, the pulsar is recycled. More formation studies and observational data would be needed to derive population characteristics for neutron star-planet systems. We focus here on the other binaries mentioned above, that are more documented.

\subsection{The octupolar regime dominates over most of the binary parameter space}
Figure~\ref{fig:epsilon} shows the values of the octupolar efficiency term $\epsilon$, depending on the companion orbital elements. Each type of companion (white dwarf, black hole or neutron star) covers a different region of the allowed parameter space.

Interestingly, one can see that most systems will be found in the region where  $\epsilon=0.1-10^{-4}$, dominated by octupolar dynamics. Therefore, we concentrate in the following on the octupolar regime and derive our main estimates within these dynamics (the full derivation for the quadrupolar regime can be found in the Appendix~\ref{app:quadrupolar}). For systems approaching $\epsilon=10^{-4}$ (NSWD systems in particular), the quadrupolar dynamics will start to dominate. However, our results should be equally valid in this case. As we demonstrate in the Appendix~\ref{app:quadrupolar}, the quadrupolar regime leads to a less efficient Kozai-Lidov mechanism, and hence to lower FRB rates per source. This is nevertheless compensated by a higher source population rate for NSWD (see Table~\ref{table:DNSchar}).

We notice that the time-scales of the octupolar regime and quadrupolar regime (see Equation~\ref{eq:dtEKL}) only  differ by a factor $128 \sqrt{10}/\qty(15\pi \sqrt{\epsilon}) \sim 8.6 \, \epsilon^{-1/2}$. Therefore in principle, systems in the octupolar regime should be characterized with longer time-scales than in the quadrupolar regime.
From Figure~\ref{fig:tEKL}, we can see that for almost all systems, the absolute time-scale of Eccentric Kozai-Lidov oscillations is longer than one year, except for very close or highly eccentric systems. Regarding the relative time delays of Eccentric Kozai-Lidov oscillations, Figure~\ref{fig:dtEKL} shows typical time-scales below one day up to times longer than the age of the Universe.

As illustrated in Figs.~\ref{fig:tEKL} and~\ref{fig:dtEKL}, the EKM can occur over a large range of time-cales. This flexibility makes this process a good candidate to explain the diversity of observed FRB rates.

\subsection{Contributions of wide and close populations to FRBs and FRB repeaters}\label{section:wide_close}
It appears from Equation~(\ref{eq:dtEKL}) and Figs.~\ref{fig:tEKL}-\ref{fig:dtEKL} that the main parameter governing the infall rate via Kozai-Lidov effect is the orbital separation between the neutron star and the black hole $a_{\rm c}$. The distance at which the neutron star binary companion can be located spans several orders of magnitude, from $a_{\rm c}\sim$ few $10^{-3}\,$A.U. to 100\,A.U.. From the previous Section, systems can be split into three populations: wide systems with $a_{\rm c} \gtrsim 10\,$A.U., mildly close systems with $a_{\rm c}\sim 0.3-{\rm few}\,$A.U., and close systems with $a_{\rm c}\lesssim 0.3\,$A.U.. 

The close systems are often associated to recycled pulsars, which are not magnetized enough to produce FRB emission at the Jansky level, except for extremely large asteroids (see Section~\ref{section:pulsar_param} and Equation~\ref{eq:FRB_flux}). 

The time-scale over which the Kozai-Lidov effects can take place, hence the lifetime of the system as an FRB source, is highly dependent on $a_{\rm c}$ (Equation~\ref{eq:tEKM}). While wide binaries have $t_{\rm EKL}\gg 10\,$yrs (or $t_{\rm KL}\gg 10\,$yrs in quadrupolar regime) and can be viewed as long-lived FRB sources, close and mildly close binaries have $t_{\rm EKL}< {\rm few}\,10{\rm s}\,$of yrs and should be considered as short-lived FRB transients. For close binaries with $a_{\rm c}\ll 1\,$A.U., $t_{\rm EKL}\ll 1\,$yr, leading to a "single-shot" transient, that will not be observed as repeating over a long time-scale. Some mildly close binaries can live thousands of years, as can be seen in Fig.~\ref{fig:tEKL} for some NSBH systems.

Close, mildly close and wide binaries are expected to be observed as different types of FRB sources for our model. Indeed, for wide systems with $a_{\rm c}\gtrsim {\rm few}\,$A.U., $\Delta t_{\rm KL}\gtrsim 10\,$yrs, leading to non-repeating sources. For close and mildly close binaries with $a_{\rm c}\lesssim {\rm few}\,$A.U., $\Delta t_{\rm KL}\lesssim 10\,$yrs, sources could be observed as repeating, with various emission frequencies. Close systems with $a_{\rm c}\ll 1\,$A.U. will produce emissions with periods shorter than a day. As previously discussed, these systems are however likely to be too faint to produce the observed FRB signals.

As the gravitational-wave merger time-scale is 
\begin{align}
t_{\rm GW}\sim& 6\times 10^{14}\, {\rm yr}\, [(M_{\rm NS}+M_{\rm c})/10\, M_\odot]^{-3} \nonumber
\\ &\times (a_{\rm c}/10\,{\rm A.U.})^4(1-e_{\rm c}^2)^{7/2} \ ,
\end{align}
the survival of both mildly close and wide binary systems over the age of the Universe is mostly guaranteed for a circular orbit ($e_{\rm c} = 0$). For large eccentricity, the merger can however happen on a shorter time-scale, down to $\sim 10^4$ years \citep{Peters64}. In any case, these time-scales are longer than $t_{\rm EKL}$ and do not need to be considered here.

Therefore, in this scenario, mildly close binaries would produce day and month-repeaters and wide binaries non-repeaters. It is interesting to notice that in the current analysis \citep{2019Natur.566..235C,Fonseca_2020}, day to few day periods seem to be favored among repeaters. This could be consistent with the dichotomy between the signatures from mildly close and wide binaries. \\

This dichotomy is reflected in the calculation of the FRB rates from these two categories. 

Mildly close binaries can be day/month-repeater FRBs during $t_{\rm KL}< 10\,$yrs. Their FRB rate density is hence directly linked to their birth rates as documented in Table~\ref{table:DNSchar}. 
We calculate roughly the total density rates $\dot{n}_{\rm c}$ for each population using the Galactic birth rates estimated in the literature and assuming a local density of galaxies of $n_{\rm gal} = 0.02$\,Mpc$^{-3}$:  $\dot{n}_{\rm c} = \dot{\nu}_{\rm c}n_{\rm gal}$.
The rate density of day-repeater FRB sources then reads
\begin{align}\label{eq:FRBrep}
\dot{n}_{\rm FRB,rep} \sim 200\,{\rm Gpc^{-3}\,yr^{-1}}  \frac{\epsilon_{\rm rep}\epsilon_{\rm mild-close}\,\dot{n}_{\rm c}}{0.2\,{\rm Mpc^{-3}\,Myr^{-1}}}\ ,
\end{align}
where $\epsilon_{\rm rep}<1$ is a source efficiency factor, and $\epsilon_{\rm mild-close}$ the fraction of mildly close systems among a population.

For wide binaries, the rate density of FRBs expected to be sourced by infalling asteroids can be estimated by convolving the mean infall rate $1/\langle\Delta  t_{\rm KL}\rangle$, the typical lifespan of the asteroid belt in its primordial configuration, $t_{\rm EKL}$ (or $t_{\rm KL}$), and the rate density of wide binary systems $\epsilon_{\rm wide}n_{\rm c}$, with $\epsilon_{\rm wide}$ the fraction of wide systems among a population. It yields a rate density of apparently non-repeating FRB events of
\begin{align}\label{eq:FRBrate}
\dot{n}_{\rm FRB,nrep} \sim& \frac{t_{\rm EKL}}{\langle\Delta t_{\rm EKL}\rangle}{\epsilon_{\rm nrep}\,\epsilon_{\rm wide}\, \dot{n}_{\rm c}} \nonumber
\\ &= \frac{1}{2}\,N_{\rm ast} \,\varepsilon_{\rm ast}^{-1}\epsilon_{\rm eff} {\epsilon_{\rm nrep}\,\epsilon_{\rm wide}\, \dot{n}_{\rm c}} \nonumber
\\ &\sim 4\times 10^3 \,{\rm Gpc^{-3}\,\rm yr^{-1}}\,\frac{N_{\rm ast}}{100}\nonumber\\
&\times \frac{\epsilon_{\rm eff}}{0.2}\left(\frac{\varepsilon_{\rm ast}}{0.15}\right)^{-1}\,\frac{\epsilon_{\rm wide}}{0.3}\, \frac{\epsilon_{\rm nrep}\, \dot{n}_{\rm c}}{0.2\,{\rm Mpc^{-3}\,Myr^{-1}}}\ ,
\end{align}
with $\epsilon_{\rm wide}=0.3$ \citep{Portegies96,Portegies_Zwart_1999,Kruckow2018}, $\epsilon_{\rm eff}\sim 0.2$ the Kozai-Lidov efficiency factor discussed in Section~\ref{section:asteroid_infall_rates}, and $\epsilon_{\rm nrep}<1$ a similar source efficiency factor as in Equation~(\ref{eq:FRBrep}). 

These calculations assume that these binaries undergo a flat source emissivity evolution, out to redshift $z\sim1$~\citep{2014LRR....17....3P}. For a star-formation type evolution, the number of sources would increase by a factor of $\sim 2$. 

In Table~\ref{table:DNSchar}, we estimated the FRB rate densities produced by various binary populations, for mildly close and wide systems, leading to repeating (rep.) and apparently non-repeating (nrep.) sources respectively. For DNS and NSWD, we have assumed that a fraction $\epsilon_{\rm mild-close}=1/3$ of the whole population was in mildly close orbit. For NSMS and NSBH, we assumed that the majority of the population was in mildly close orbit $\epsilon_{\rm mild-close}=1$. For NSWD, we assumed that $\epsilon_{\rm wide}=1/3$, and used the rates provided in the literature for wide NSMS and DNS rates. For NSBH, we assumed $\epsilon_{\rm wide}=0.1$. These fractions are discussed in Section~\ref{section:binary_char}.\\

The rate densities estimated from Equation~(\ref{eq:FRBrate}) can be directly compared to the cosmological FRB rate densities inferred from observations, of order $\dot{n}_{\rm FRB,obs}\sim 2\times 10^3 \,{\rm Gpc}^{-3}\, {\rm yr}^{-1}$ \cite{Petroff_2019}. Except for NSBH, for which the entire population would not suffice to produce the observed FRB rate densities, we notice that $\dot{n}_{\rm FRB,obs}\ll \dot{n}_{\rm FRB,nrep}$. The inferred source efficiency can thus be of order $\epsilon_{\rm nrep}\lesssim  10\,\%$ (NSWD: 2.0\%, NSMS: 8.7\%, DNS: 7.4\%). This number leaves room for binary systems which do not fulfill the criteria to undergo Kozai-Lidov mechanisms, such as systems without asteroid belts, orbital inclinations, etc. 

More than 700 FRBs have been observed as of today, (although only 137 have been published), among which \valentin{22} have been identified as repeaters \citep{collaboration2019chimefrb,Fonseca_2020}, yielding a possible ratio of $\sim 3\%$. Interestingly, this number match quite well the ratios estimated for our systems:
$\dot{n}_{\rm FRB,rep}/\dot{n}_{\rm FRB,nrep}\sim 5.3\%\,\epsilon_{\rm rep}/\epsilon_{\rm non-rep}$ for NSMS (for NSMS: 1.4\% and for DNS: 1.4\%). For a same population, one can assume that $\epsilon_{\rm rep}=\epsilon_{\rm non-rep}$. However, the formation of asteroid belts might differ for close and wide systems. 

It is possible that all the neutron-star binaries mentioned above contribute to the FRB rates. If one assumes that their efficiencies $\epsilon_{\rm nrep}$ and $\epsilon_{\rm rep}$ are equal, the total rate density of apparently non-repeating FRBs would be of $\dot{n}_{\rm FRB,nrep,all}\sim 15\times 10^4 \epsilon_{\rm nrep}\,{\rm Gpc}^{-3}\, {\rm yr}^{-1}$, with a source efficiency that can be as low as $\epsilon_{\rm nrep}=1.4\%$. The repeater rate density would be of $\dot{n}_{\rm FRB,nrep,all}\sim 3.1\times 10^3 \epsilon_{\rm nrep}\,{\rm Gpc}^{-3}\, {\rm yr}^{-1}$, which implies a repeater to non-repeater ratio of $\sim 2\%$, compatible with the observed ratios. The scenario is surprisingly comfortable and consistent with the current observations.
\begin{figure}[tb!]
	\center
	\includegraphics[width = 0.99\linewidth]{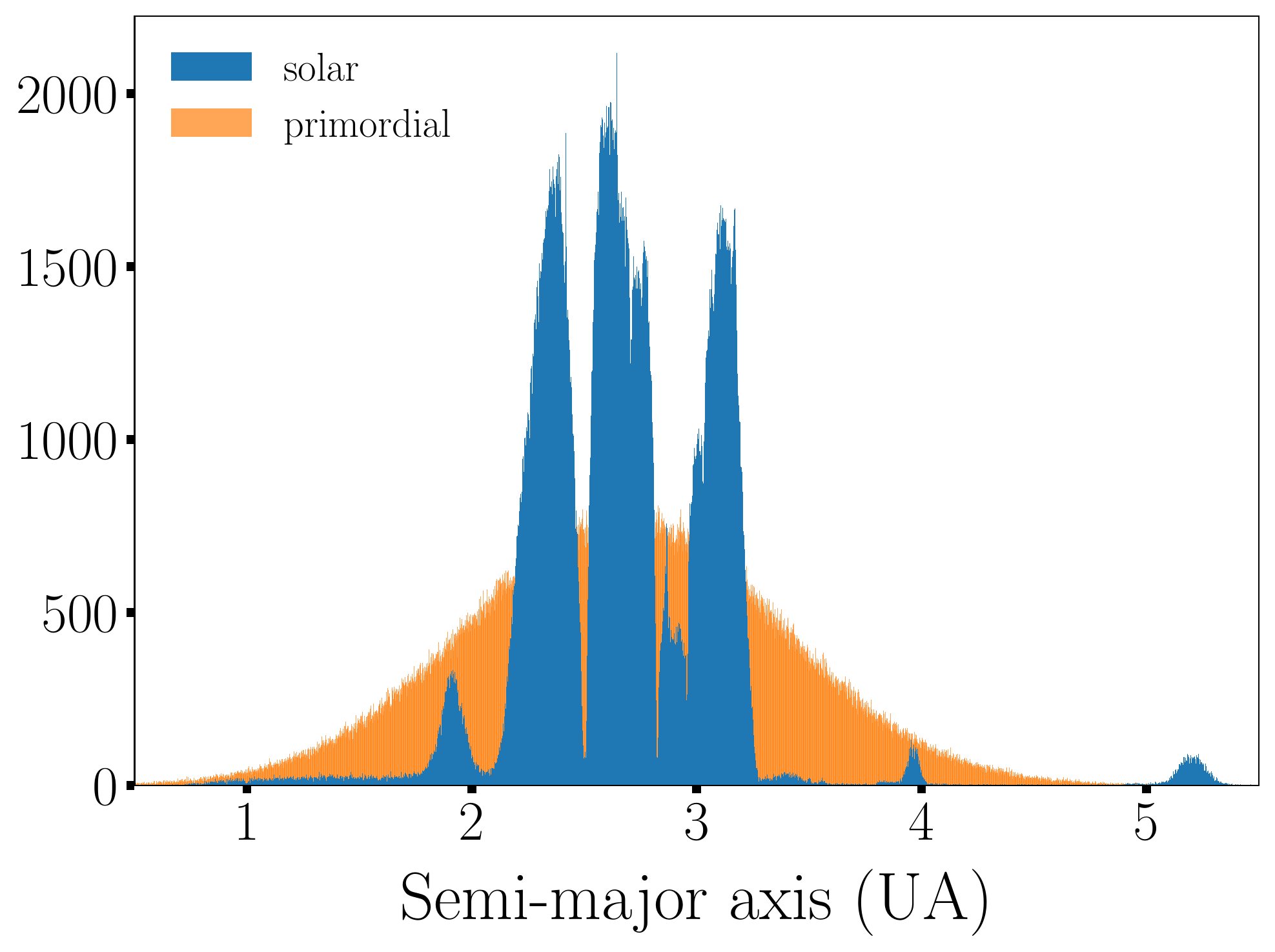}
	\includegraphics[width = 0.99\linewidth]{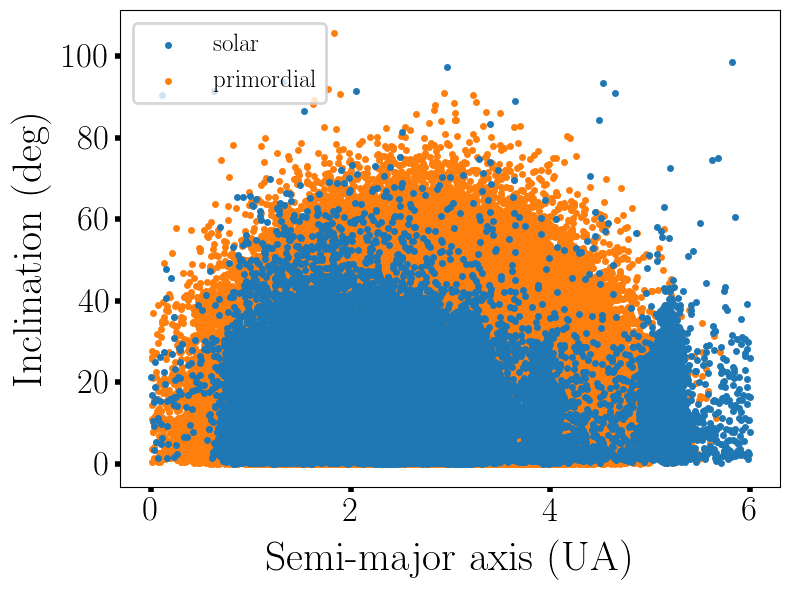}
	\caption{Asteroid orbital parameter distributions inside the Solar asteroid belt (blue) and our reconstructed primordial model belt (orange). {\it Top:} Asteroid semi-major axes distribution. {\it Bottom:} Asteroid inclinations as a function of the semi-major axes. In both panels, the Kirkwood gaps are clearly visible in the Solar asteroid belt.}
	\label{fig:asteroid_distrib}
\end{figure}

\section{Simulating numerically asteroid infall rates} \label{section:application_belt}
In this section, we simulate numerically the FRB rates of close and wide binary systems with an asteroid belt undergoing Kozai-Lidov effects. We model a primordial asteroid belt (without any gaps such as the Kirkwood gaps of the Solar system), in analogy with the Solar asteroid belt.

\subsection{Synthetic asteroid belt}
We model the distribution of the orbital parameters of the current Solar belt using the data from the IAU Minor Planet Center~\citep{MPCOD}. A total number of $792041$ asteroids of the Solar belt are inventoried in this database.

Numerous asteroids sensitive to the Kozai-Lidov effect are missing from the distribution of orbital elements of the current Solar asteroid belt, influenced by giant planets such as Jupiter. The Kirkwood gaps for instance, illustrate this effect. 
These features motivate the construction of a synthetic asteroid belt for our model, filling most of the gaps and mimicking the primordial population of the belt (see Figure~\ref{fig:asteroid_distrib}).

The synthetic belts follow a gaussian distribution fitting the general trend of the current Solar belt.
We use for the semi-major axis a standard deviation of $\sigma_{a} = 0.15 \langle a_{\rm ast}\rangle$, with the mean semi-major axis $\langle a_{\rm ast}\rangle$ left as a free parameter. For the inclinations, we follow the Solar belt distribution with mean inclination $\langle i_{\rm ast}\rangle = 0^{\circ}$ and standard deviation $\sigma_{i} = 30^{\circ}$. 
The initial eccentricities are not modeled since they are not relevant to our computation of the Kozai-Lidov effects. However Figure~\ref{fig:asteroid_distrib} suggests that a fitting model similar to what is done for the distribution of semi-major axis could be easily achieved.

The number of these massive asteroids follows a power law distribution as a function of their size, as observed in the Solar System \cite{MPCOD} (see Section~\ref{section:asteroid_size}. Their masses can be retrieved by assuming that they are roughly spherical, with a density $\rho_{\rm ast}=2\,{\rm g\,cm}^{-2}$.

This simple method allows us to construct a more generic asteroid belt, although it is restricted to our knowledge of the Solar system. 

\begin{figure}[tb!]
	\center
	\includegraphics[width = 0.99\columnwidth]{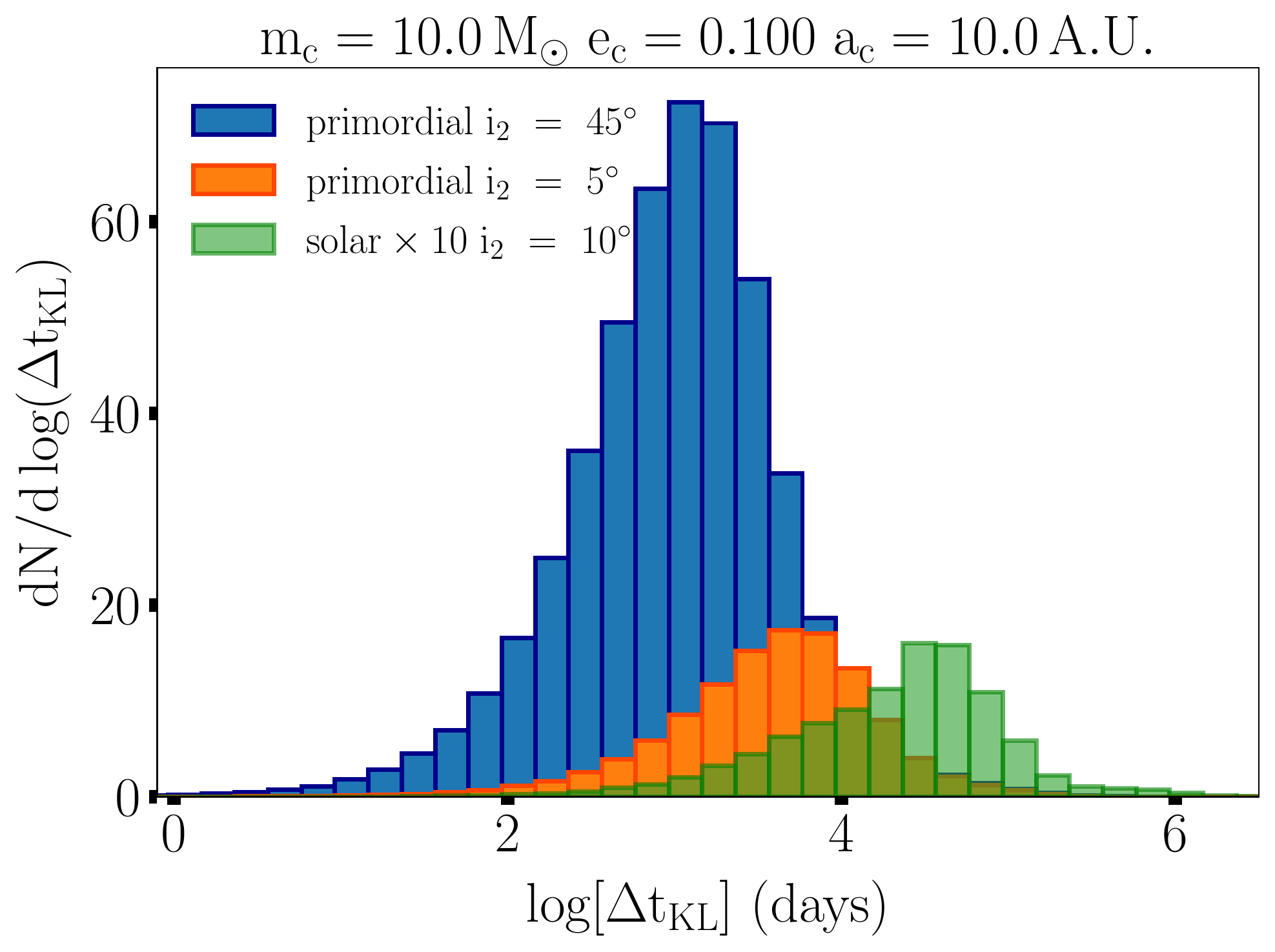}
	\caption{Distribution of relative time differences $\Delta t_{\rm KL}$ of  asteroids falling into the Roche lobe of the central compact object due to Kozai-Lidov oscillations (which can be directly interpreted as the FRB emission periods), for the current Solar asteroid belt (green) and the primordial belt, for an inclination of the outer companion plane $i_{\rm c}=5\degree$ (orange), $i_c=10\degree$ (green) and $i_{c}=45\degree$ (blue), and initial asteroid number $N_{\rm ast}=10^2$. We consider a high mass $M_c=10\,M\odot$ wide system with $a_{\rm c} = 10$\,A.U. and $\expval{a_{\rm ast}}= 1$\,A.U.. The number of asteroid infalls increase with the inclination of the system, as expected. Additionally, the relative time differences $\Delta t_{\rm KL}$ increase with lower inclinations, except for the Solar asteroid case. The present Solar asteroid belt has already undergone Kozai-Lidov effects during its lifetime, leading to a cleansing of its asteroids that is potentially sensitive to Kozai-Lidov oscillations, which explains its misleading behavior in this figure.}
	\label{fig:rates_inclinations}
\end{figure}
\begin{figure*}[tb!]
	\center
	\includegraphics[width = 0.49\linewidth]{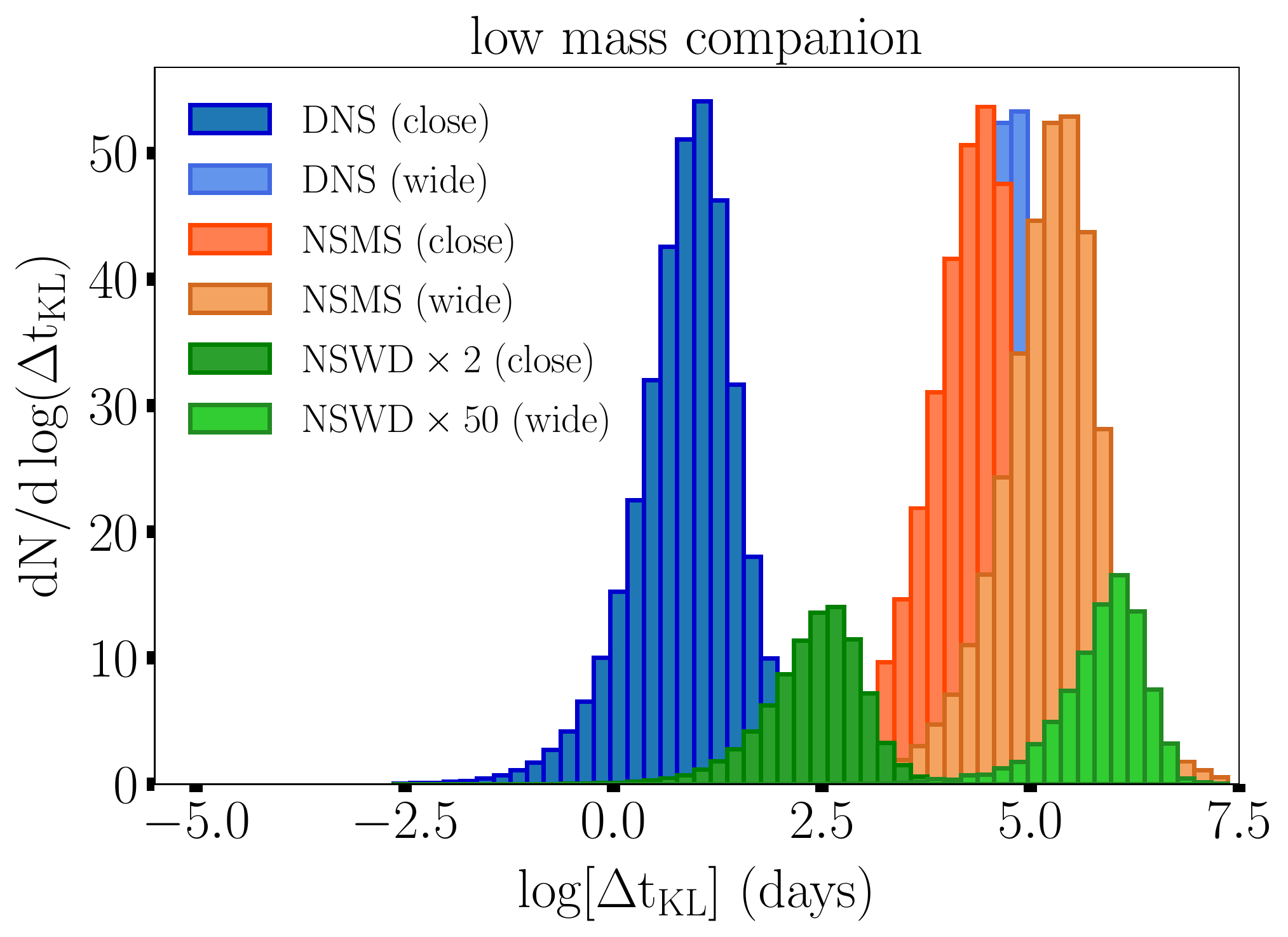}
	\includegraphics[width = 0.49\linewidth]{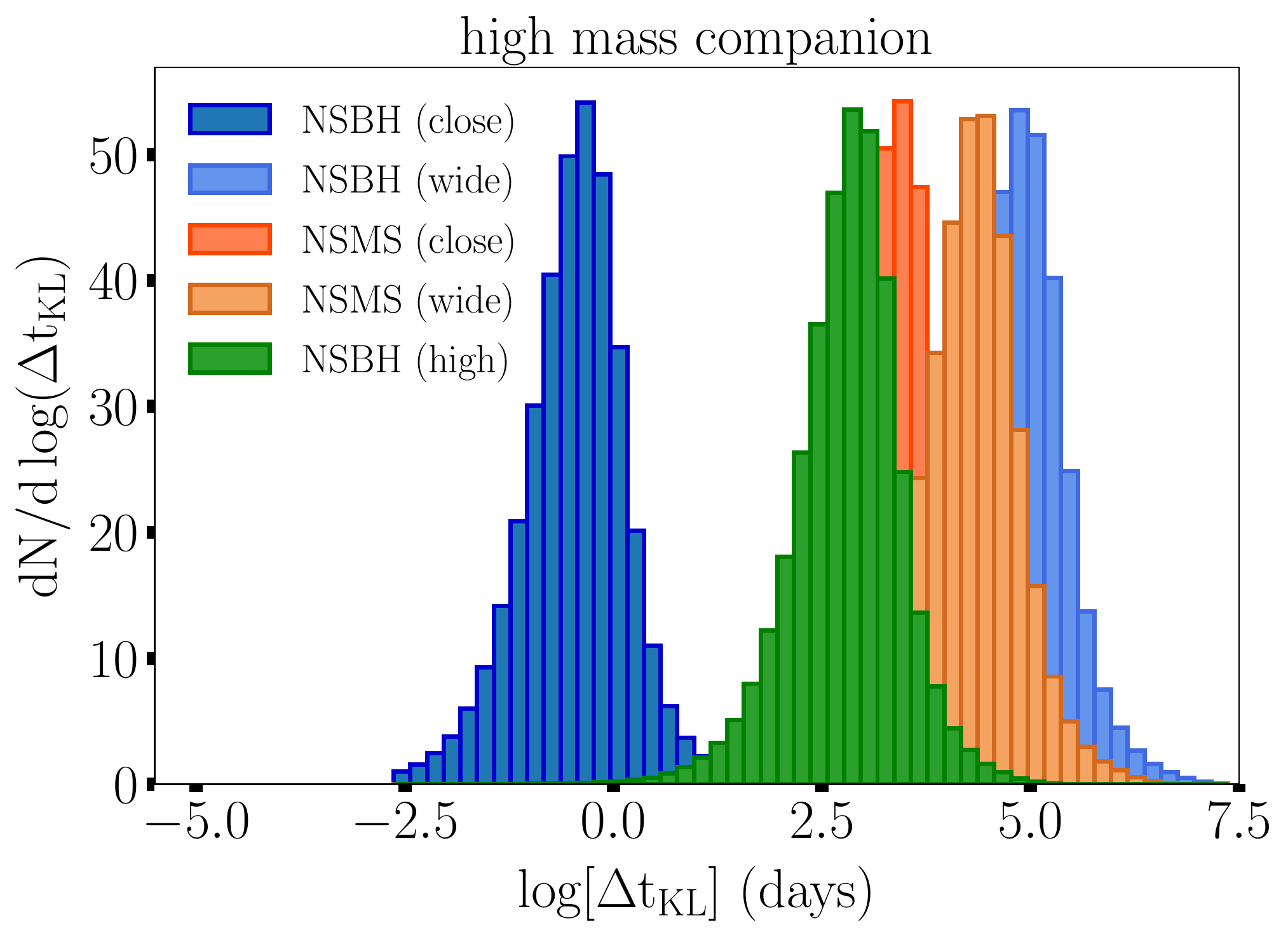}
	\caption{Same as Figure~\ref{fig:rates_inclinations}, distribution of relative time differences $\Delta t_{\rm KL}$ of asteroids falling into the Roche lobe of the central compact object due to Kozai-Lidov oscillations, for the different systems considered here. {\it Left:} low mass companions. {\it Right:} high mass companions. The initial number of asteroids is $N_{\rm ast}=10^2$ and the inclinations of the companions is $i_c=30\degree$. Specific parameters for each configuration can be found in Table~\ref{table:simu_config}.}
	\label{fig:rates_light_heavy}
\end{figure*}
\begin{table}
\begin{center}
\begin{tabular}{lrrrrr}
\hline
\hline
{\bf Systems} & \begin{tabular}{@{}r@{}} $M_c$\\ $\qty[M_\odot]$\end{tabular} & $e_c$ & \begin{tabular}{@{}r@{}} $a_c$ \\$\qty[\rm A.U.]$\end{tabular} & \begin{tabular}{@{}r@{}} $\langle a_{\rm ast}\rangle $\\ $\qty[\rm A.U.]$\end{tabular} \\
\hline
DNS (close)        & $1.4$ & $0.01$    & $0.1$ & $0.02$ \\
DNS (wide)         & $1.4$ & $0.7$     & $30$  & $1$    \\
\hline
NSMS (low, close)  & $1$   & $0.6$     & $5$   & $0.1$  \\
NSMS (low, wide)   & $1$   & $0.8$     & $50$  & $1$    \\
\hline
NSWD (close)       & $0.1$ & $10^{-5}$ & $1$   & $0.1$  \\
NSWD (wide)        & $0.1$ & $0.01$    & $30$  & $1$    \\
\hline
NSBH (close)       & $10$  & $0.1$     & $0.1$ & $0.02$ \\
NSBH (wide)        & $10$  & $0.1$     & $30$  & $1$    \\
\hline
NSMS (high, close) & $10$  & $0.6$     & $5$   & $0.1$  \\
NSMS (high, wide)  & $10$  & $0.8$     & $50$  & $1$    \\
\hline
NSBH (high)        & $100$ & $0.7$     & $30$  & $1$    \\
\hline
\hline
\end{tabular}
\caption{Configuration parameters for the Monte-Carlo simulations presented in Section~\ref{section:application_belt} and results in Figure~\ref{fig:rates_light_heavy}, for low and high mass systems, and close and wide companion orbits.}
\label{table:simu_config}
\end{center}
\end{table}

\subsection{Simulations set-up} \label{section:simulation_setup}
Following the asteroid distribution computed in the previous section, we randomly draw a set of asteroid parameters (size $R_{\rm ast}$, semi-major axis $a_{\rm ast}$, inclination $i_{\rm ast}$). We select the objects that meet the following three criteria
\begin{enumerate}
    \item minimum size $R_{\rm ast}> 50$\,km, large enough to trigger FRB-like emissions via the Alfv\'en mechanism (Equation~\ref{eq:FRB_flux})
    
    \item allow the triggering of Kozai-Lidov oscillations (see Section~\ref{app:quadrupolar})
    
    \item can reach the Roche limit under the Kozai-Lidov effect (Equation~\ref{eq:eta}).
\end{enumerate}
The last two criteria are always met under the octupolar regime. Both quadrupolar and octupolar regimes are taken into account in this calculation, as well as the GR effects. 

For the selected asteroids, we compute the Kozai-Lidov timescales needed to reach the maximal eccentricity and the relative time delays between two consecutive asteroid infalls. The distribution of these relative infall times can be directly compared to FRB rates. To avoid statistical fluctuations due to the Monte-Carlo drawing, we average our results over $10^{4}$ simulations.

\subsection{Asteroid infall rates for a Solar-like belt} \label{section:asteroid_infall_rates}
Figure~\ref{fig:rates_inclinations} shows the distribution of the relative time delays $\Delta t_{\rm KL}$ for asteroids falling onto the central neutron star, for a current (green) and primordial Solar-like belt, and for companion inclinations $i_{\rm c}=5\degree$ (orange), $i_{\rm c}=10\degree$ (green) and $i_{\rm c}=45\degree$ (blue). The central neutron star has mass $M_{\rm NS} = 1.4\,M_{\odot}$ and the outer companion $M_{\rm c} = 10\,M_{\odot}$. The initial number of asteroids is set to $N_{\rm ast}=10^2$, following the power law spectra observed in the Solar system belt for the most massive asteroids.
We examine in Figure~\ref{fig:rates_inclinations} the effect of the companion inclination $i_c$ on the relative timescales and efficiency of the Kozai-Lidov effect, in the case of a wide system with companion distance $a_{\rm c}=10$\,A.U. and mean asteroid belt distance $\expval{a_{\rm ast}}=1$\,A.U..

For this wide system, the infall rates span from days to thousands of years, with a maximum around $\langle \Delta t_{\rm KL} \rangle\sim 10-100$\,years, depending on the inclination $i_2$. More interestingly the efficiency of the Kozai-Lidov process, the ratio of the number of falling asteroids over the number of drawn asteroids, is greater for inclined systems. More asteroids fall onto the central pulsar for a more inclined asteroid belt, which is consistent with the Kozai-Lidov process, since more asteroids will meet the Kozai-Lidov criterion on the inclination $i_{\rm ast}\gtrsim 40\degree$. Nevertheless, these simulations show that even for low to mildly inclined systems, the efficiency remains around $20\%$.

The comparison between the current Solar belt (green) and the primordial belts (blue or orange) shows that the lack of Kirkwood gaps induces a drastic increase of short time-scales in the asteroid infall rate, and depending on the inclination, a factor of a few to an order of magnitude more events in total (greater efficiency).

Figure~\ref{fig:rates_light_heavy} displays, similarly to Figure~\ref{fig:rates_inclinations}, the distribution of the relative time delays $\Delta t_{\rm KL}$ for asteroids falling onto the central neutron star, for various neutron star systems. The left panel presents systems with low mass companions, such as DNS, NSMS and NSWD for close and wide systems. The right panel shows systems with high mass companions, namely NSBH and NSMS. Wide and close systems can be distinguished through their relative asteroid falling rates: wide systems induce higher rates than close systems, for both low and high mass companions. NSWD systems appears to be much less efficient than any other systems, this is due to the fact that the Kozai-Lidov time-scale is a function of the mass of the companion (see Eqs.~\ref{eq:dtKL} and~\ref{eq:dtEKL}). An opposite result can be seen for high mass NSBH systems, which are much more efficient and with shorter time-scales than NSBH wide systems, as expected. Finally one can notice that most of the systems (except NSWD systems), present an efficiency above $50\%$ in this Kozai-Lidov mechanisms.

\subsection{Connection with FRB observations} \label{section:connectionFRB}

The results of the simulations detailed in Section~\ref{section:asteroid_infall_rates} show a high consistency with the analytical estimates computed in Sections~\ref{section:KL_time} and~\ref{section:wide_close}. These simulations demonstrate that the application of the Kozai-Lidov framework introduced in Section~\ref{section:secular_dynamics} to a multiplicity of small objects such as the ones found in the Solar asteroid belt, remains consistent with the conclusions drawn in Section\ref{section:wide_close}. They validate that the various populations of binary pulsar systems, such as DNS, NSMS, NSWD and NSBH, can explain the dichotomy observed between repeating and non-repeating FRBs.

The efficiency of the Kozai-Lidov process is illustrated on Figure~\ref{fig:rates_light_heavy}, where the number of asteroid infalls compared to the total number of asteroid simulated ($N_{\rm ast}=100$ in Figure~\ref{fig:rates_light_heavy}), corresponds to the efficiency of the Kozai-Lidov mechanism in driving asteroids down to the Roche limit (the factor $\epsilon_{\rm eff}$ introduced in Section~\ref{section:DtKL}, see also Section~\ref{section:simulation_setup}). It is clear that for most pulsar binary systems this process is efficient with ratio largely above $50\%$, and even for NSWD systems which are the less efficient systems, this ratio is around $20\%$. Consequently the Kozai-Lidov process in pulsar binary systems is efficient in driving asteroids down to the Roche limit from a Solar-like asteroid belt in our model.

Another interesting result coming only from the simulations concerns the distribution tails displayed in Figure~\ref{fig:rates_light_heavy}. One can see that the Gaussian rates distributions (at a first approximation) possess extended distribution tails. This result implies that for some wide systems, with long time delays on average, events could occur with shorter time delays at some point in the process. This translates, in terms of FRB bursts, in the existence of some observed non-repeating sources that can produce few repetition bursts once in a while. One should note that these repetitions would be highly irregular, and would not be sustained over time, as they are statistically rare.

This spread in the time delays is also valid for close systems, with short time delays, and associated with FRB repeaters. This numerical result is actually in agreement with observations since a fraction of FRB sources are found bursting with irregular short periods, ranging from days to month-time-scales. These bursts would correspond to the left-hand tail of distributions such as the one shown in Figure~\ref{fig:rates_light_heavy} for close systems. 

FRB121102 could be a good candidate for this tail scenario. Activity periods have been reported for hour scale periods, day scale periods and monthly periods (Table 2 in~\cite{Scholz_2016}). Such an erratic behavior could well be explained as a tail of the asteroid falling rate distribution. This source also presents substructure in the signal, with fainter pulses arriving at shorter intervals \citep{Zhang2018}. These could be explained by the fragmentation of asteroid during the disruption in the Roche lobe, as mentioned in Section~\ref{section:beam} or simply the presence of asteroids clumping in the asteroid belt, as observed in the Solar system, which is explained by asteroid collisions leading to subgroups of asteroids close-by and with similar orbital paramaters, therefore leading to similar Kozai-Lidov time delays and so similar infall rates.

The close systems presented in Figure~\ref{fig:rates_light_heavy} illustrate the possibility of having a population of short-lived repeaters, with day-scale periods. These sources will appear less numerous than the wide systems due to their short active timescale (see Figure~\ref{fig:tEKL}), which is consistent with the low percentage of repeaters observed so far.

Finally the existence of short transient sources is predicted with our model. From Figure~\ref{fig:tEKL}, it is possible to find sources with very short lifetimes, below one year. These sources would completely deplete their asteroid belt over very brief infall rates, resulting in a firework display of bursts. These close sources are associated with recycled pulsars, with magnetic fields that are too low to produce Jansky level bursts. These events should hence be difficult to observe because of their brevity and their low flux.

\section{Conclusion and discussion}\label{section:discussion}
Fast radio bursts can be produced if asteroids pass close to the Roche limit of a compact object with an electromagnetic wind \citep{Mottez2014,mottez2020repeating}, or if they undergo  collisions with this object \citep{Dai_2016, Smallwood_2019}. The infall of asteroids from standard belts onto the central compact object can be triggered by Kozai-Lidov oscillations, in the presence of an outer black hole. 

The asteroid dynamics described by our model is able to reproduce the overall observed ratio of repeating to non-repeating FRBs and motivates an explanation to unify the two observed populations under one simple mechanism, already evidenced in the Solar system. FRBs could be comfortably produced by a population of neutron star binary systems, in particular by NSWD, NSMS and DNS binaries. NSBH systems are expected to have a lower contribution due to their lower population rates. We find that mildly close systems (companion semi-major axis $a_{\rm c}\sim 0.3-{\rm few}\,$A.U.) produce day/month scale repeaters that live $<10\,$yrs, while wide systems ($a_{\rm c}\sim {\rm few}-10{\rm s}\,$A.U.) are steady sources, which will be observed as non-repeating.

We find that a comfortable fraction of a few percent ($<10\%$) of these binary systems in the Universe can account for the observed non-repeating FRB rates. More remarkably, our wide/close orbit dichotomy model predicts a ratio between repeating and non-repeating sources of a few percent, which is in good agreement with the observations.

Close systems with $a_{\rm c}\ll 1\,$A.U. could also lead to beamed radio signals, but such systems being often associated with recycled pulsars with low magnetic fields, the FRB flux should be low. The signatures of such systems would be specific: a series of mJy level pulses arriving over seconds to hours, and that would never repeat again. 
Sub-Jansky radio bursts arriving with short periods ($\ll\,$day) produced in a single shot could thus constitute an electromagnetic counterpart to NSWD, DNS and NSMS mergers. Such FRBs could also be a counterpart to NSBH mergers as was already predicted in~\citep{Kotera_Silk16}.

Simulations presented in Section~\ref{section:application_belt} numerically validate  the analytical conclusions drawn in section~\ref{section:FRBrates}. We find that our conclusions hold under more realistic conditions, for instance when taking into account a realistic distribution of asteroid parameters inside an asteroid belt. Finally, the simulations also show that the asteroid belt structure combined with the induced dynamics of specific pulsar systems can lead to a short time-scale tail (or repetition tail) even for systems labeled as non-repeaters.

Three major predictions can be made from our scenario, which can be tested in the coming years:
\begin{enumerate}
    \item Most repeaters should stop repeating after $t_{\rm EKL}<{\rm few}\,10{\rm s}$ of years, as their asteroid belts becomes depleted.
    
    \item Some non-repeaters could occasionally repeat, if we hit the short $\Delta t_{\rm EKL}$ tail of the FRB period distribution.
    
    \item Series of sub-Jansky level short radio bursts could be observed as electromagnetic counterparts of NSWD, DNS, NSMS and NSBH mergers.
\end{enumerate}

The present study can be applied to other close binary systems, provided that the central object generates a magnetized wind. In particular, pulsar systems with planets could contribute to this scenario. 

The recent observation of two intense radio bursts in coincidence with X-ray flares (\citealp{2020Natur.587...59B,collaboration2020bright,mereghetti2020integral}), expected to originate from the magnetar SGR1935+2154, has shown some similarities with FRB emissions. This observation, if attributed to an FRB-like signal, would be the first FRB event observed in our Galaxy but also the dimmest FRB ever observed, with 40 times less radiated energy. Our model is not incompatible with this observation, assuming that this magnetar is in a binary configuration (even with a very far away companion). Some dynamical configurations, resulting from Kozai-Lidov oscillations can result in the observation of a double radio burst: (i) the observation of two consecutive (and close-by) asteroids falling close to the Roche limit and radiating via the Alfv\'en wing mechanism, (ii) the observation of a single asteroid close to the Roche limit but observed twice thanks to the turbulence of the beam, crossing twice the line-of-sight of the observer, (iii) the fortuitous observation of the disruption of an asteroid crossing the Roche limit and emitting multiple radio beams in random directions, and crossing twice the line-of-sight of the observer. However the production of the coincident X-ray flares remains more challenging. 

One possibility relies in the accretion of tidally disrupted material from a single  asteroid onto the magnetar (e.g.,\citealp{1994ApJ...437..727K}). 

A rough estimate can be made by assuming emission via disruption and  Eddington accretion of an asteroid of size $R_{\rm ast}\sim 100$\,km and mass $M_{\rm ast}\sim 8\times 10^{24}$\,g (for a density $\rho_{\rm ast}\sim 2$\,g.cm$^{-3}$) at the Roche limit $d_{\rm Roche}$ and falling onto  the central neutron star of size $R_{\rm NS}\sim10$\,km and mass $M_{\rm NS}\sim 1.4\,M_\odot$. The mass accretion rate can be estimated as $\dot{M}_{\rm ast} = \epsilon_{\rm ast} M_{\rm ast} / t_{\rm fall}$, where $\epsilon_{\rm ast} \sim 0.1$ is the fraction of asteroid material accreted and $t_{\rm fall} = \sqrt{2 d_{\rm Roche}/GM_{\rm NS}}\sim 1.5\,$h the infall time from the Roche limit down to the neutron star. The Eddington luminosity is hence given by $L_{\rm edd, ast} = \epsilon_{\rm edd} \dot{M}_{\rm ast} c^2 \sim 10^{36}\,$erg.s$^{-1}$, where $\epsilon_{\rm edd}\sim 10^{-5}$ represents the efficiency of the Eddington process (expected to be much less efficient than for  black hole accretion). The corresponding isotropic equivalent energy is $E_{\rm iso, ast}\sim 10^{39}\,$erg, close to the value inferred from observations ($E_{\rm iso, obs}\sim 1.4 \times 10^{39}\,$erg, \citealp{mereghetti2020integral}). Finally, if the emission results from thermal processes, the effective blackbody temperature $T_{\rm eff, ast}$ can be obtained through the Stefan-Boltzman law, which enables the determination of the maximal photon energy $E_{\rm \gamma, ast}=h \nu \sim k_{\rm B} T_{\rm eff}\sim 204\,$keV also close to the observations ($E_{\rm gamma, obs}\sim 20-200$\,keV, \cite{mereghetti2020integral}).

Alternatively the interactions of material from the plasma with the Alfv\'en wings could also lead to high energy photon emission, for instance following similar processes as suggested by~\cite{2013ApJ...762...13B}.


\section*{Acknowledgements}
We thank the anonymous referee for the thoughtful comments which helped improve this paper. We also thank F. Antonini, A. Benoit-L\'evy, G. Bou\'e, F. Daigne, R. Duque, I. Dvorkin, A. Lamberts, and P. Zarka for very fruitful discussions. This work is supported by the APACHE grant (ANR-16-CE31-0001) of the French Agence Nationale de la Recherche. This research has made use of data provided by the IAU's Minor Planet Center.

 \bibliographystyle{aa} 
 \bibliography{biblio.bib} 
 
\appendix

\section{Three-body dynamics}\label{app:threebody}
In this appendix, we shortly review the key ingredients to understand the three-body dynamics in the framework of the Kozai-Lidov mechanisms. We detail how to obtain the various Kozai-Lidov periods and time-scales for the quadrupolar and octupolar regimes. The demonstration follows a standard approach, shown by \cite{Antognini_2015,Lithwick_2011, Naoz_2016} and others.

We define the Delaunay variables in the framework of canonical angle-action variables
\begin{align}
    l_1 \to L_1 &= \frac{m_0 m_1}{m_0 + m_1} \sqrt{G \qty(m_0 + m_1) a_1}
    \\l_2 \to L_2 &= \frac{m_2 \qty(m_0 + m_1)}{m_0 + m_1 + m_2} \sqrt{G \qty(m_0 + m_1 + m_1)a_2}
    \\ g_1 \to G_1 &= L_1 j_1
    \\ g_2 \to G_2 &= L_2 j_2
    \\ h_1 \to H_1 &= G_1 \cos{\qty(i_1)}
    \\ h_2 \to H_2 &= G_2 \cos{\qty(i_2)}
\end{align}
with $j_k = \sqrt{1 - e_k^2}$. This set of variables preserves the canonical structure of the Hamiltonian description and allows to fully describe the three-body system.

\subsection{The quadrupolar regime and the Kozai-Lidov oscillations}
The quadrupolar term from Equation~\eqref{eq:Hpert} is a constant of motion since the energy is conserved at the quadrupole order. We can rewrite the quadrupolar term as follows
\begin{align}
    \mathcal{H}_{\rm quad} =& C_2 \big[ \qty(2+3e_1^2) \qty(1 - 3 \cos^2{i_{\rm tot}}) \nonumber
    \\ &- 15 e_1^2 \qty(1 - \cos^2{i_{\rm tot}}) \cos{2g_1} \big] \ ,
\end{align}
which gives a reduced Hamiltonian $\tilde{\mathcal{H}}_{\rm quad} = \mathcal{H}_{\rm quad}/ C_2$, also a constant of motion.
Furthermore we have 
\begin{align} \label{eq:momenta_conservation}
    \cos{\qty(i_{\rm tot})} = \frac{G_{\rm tot}^2 - G_1^2 - G_2^2}{2 G_1 G_2} \ ,
\end{align}
from the Al-Kashi theorem applied to the angular momenta. In the test particle limit (where $m_1 \to 0$) Equation~\eqref{eq:momenta_conservation} reduces to $G_{\rm tot} \approx G_2 + G_1 \cos{\qty(i_{\rm tot})}$, and since $G_{\rm tot}$ and $G_2$ are conserve quantities, it comes that $j_1 \cos{\qty(i_{\rm tot})}$ is therefore conserved. Usually the previous constant of motion is defined as
\begin{align} \label{eq:Kozai_integ}
    \Theta = \qty[j_1 \cos{\qty(i_{\rm tot})}]^2 \ ,
\end{align}
and called Kozai's integral. Equation~\eqref{eq:Kozai_integ} represents the projection of the total orbital momentum along the z-axis: $j_{\rm z}$, and allows to compute the transfer of inclination to eccentricities via
\begin{align}
    \sqrt{1 - e_{\rm 1, min}^2} \cos{\qty(i_{\rm tot, max})} = \sqrt{1 - e_{\rm 1, max}^2} \cos{\qty(i_{\rm tot, min})} \ .
\end{align}
So the reduced Hamiltionian can be rewritten as
\begin{align}
    \tilde{\mathcal{H}}_{\rm quad} =& j_1^{-2} \bigg[ \qty(5 - 3j_1^2) \qty(j_1^2 - 3 \Theta) \nonumber
    \\ &- 15 \qty(1 - j_1^2) \qty(j_1^2 - \Theta) \cos{\qty(2 g_1)}
     \bigg] \ .
\end{align}
Finally the reduce Hamiltonian is a function of $\tilde{\mathcal{H}}_{\rm quad} = f\qty[e_1, \cos{\qty(i_{\rm tot})}, g_1]$, hence with three degrees of freedom but since $\tilde{\mathcal{H}}_{\rm quad}$ and $\Theta$ are constant of motion the system is fully integrable via the equations of motions.

From the canonical variable $g_1$, we can write the standard equation of motion
\begin{align} \label{eq:canonical_g1}
    \dv{g_1}{t} &= \pdv{\mathcal{H}_{\rm quad}}{G_1} = \frac{C_2}{L_1} \pdv{\tilde{\mathcal{H}}_{\rm quad}}{j_1}
    \\ &= \frac{6 C_2}{L_1} j_1^{-3} \bigg[ 5 \qty(\Theta - j_1^4) \qty(1 - \cos{\qty(2 g_1)} + 4 j_1^4) \bigg] \ .
\end{align}
And since $j_1$ is related to $G_1$ via $L_1$ (constant of motion), we have
\begin{align} \label{eq:evolution_j1}
    \dv{j_1}{t} &= \frac{1}{L_1} \pdv{\mathcal{H}_{\rm quad}}{g_1} = \frac{C_2}{L_1} \pdv{\tilde{\mathcal{H}}_{\rm quad}}{g_1}
    \\ & \frac{30 C_2}{L_1} j_1^{-2} \qty(1 - j_1^2) \qty(j_1^2 - \Theta) \sin{\qty(2 g_1)} \ .
\end{align}
Finally, $\cos{\qty(i_{\rm tot})}$ can be solved thanks to Equation~\eqref{eq:evolution_j1} injected in Equation~\eqref{eq:momenta_conservation}, which provides a full set of integrable equations to describe the full dynamics of the three-body system at the quadrupole order.

\cite{Antognini_2015} has shown that thanks to the integrability of the quadrupole order, the exact Kozai-Lidov period can be derived.
\begin{align}
    t_{\rm KL} = \oint \dd{t} = \oint \dv{t}{j_1} \dd{j_1} \ ,
\end{align}
over a full period. Hence from Equation~\eqref{eq:evolution_j1}, \cite{Antognini_2015} shows that the exact period can be written as
\begin{align} \label{eq:KL_period}
    t_{\rm KL} =& \frac{L_1}{15 C_2} \int_{j_{\rm min}}^{j_{\rm max}} \qty(1 - j_1^2)^{-1} \bigg[ \qty(1 - \frac{\Theta}{j_1^2})^2  \nonumber
    \\ &- \qty(\frac{1}{5} - \frac{\Theta}{j_1^2} + \frac{4}{5} \frac{C_{\rm KL}}{1 - j_1^2}) \bigg]^{-1/2} \ , 
\end{align}
where 
\begin{align} \label{eq:KL_constant}
    C_{\rm KL} = 1/12 \qty(2 - \tilde{\mathcal{H}}_{\rm quad} - 6 \Theta) \ , 
\end{align}
a constant of motion found by Lidov in 1962, discriminating between libration regime ($C_{\rm KL} < 0$) and rotation regime ($C_{\rm KL} > 0$).

Finally, \cite{Antognini_2015} has shown that the exact period of Equation~\eqref{eq:KL_period} differs from a factor of a few from the timescale described in Equation~\eqref{eq:tKL} (and obtained via $t_{\rm KL} \sim L_1/\qty(15 C_2)$), in dynamical regimes far from any resonances (see Figure 1 from \citealp{Antognini_2015}).

\subsection{The octupolar regime and the Eccentric Kozai-Lidov Mechanism}
Even though the three-body system is not integrable at the octupolar order, \cite{Antognini_2015} shows that nevertheless it is possible to derive the exact period of the Eccentric Kozai-Lidov mechanism.

In a first step \cite{Antognini_2015} follows the analysis of~\cite{Katz_2011}, where the authors introduce the eccentricity vector
\begin{align}
    \vec{e} = e \qty(\cos{\Omega_e} \sin{i_e}, \sin{\Omega_e} \sin{i_e}, \cos{i_e}) \ ,
\end{align}
this vector is pointing towards the periapsis of the inner binary, and allows one to describe the motion of the periapsis with time.
Since $\mathcal{H}_{\rm quad}$ is constant, from Equation~\eqref{eq:KL_constant} it is possible to define another constant of motion
\begin{align} \label{eq:phi_constant}
    \Phi_{\rm quad} = C_{\rm KL} + \frac{1}{2} \Theta \ .
\end{align}
We note that at the octupolar order, $\Theta$ and $C_{\rm KL}$ are no longer constants of motion. However, it is possible to assume that $\Theta$ and $C_{\rm KL}$ remain approximately constant over timescales of single KL oscillations. This motivates the re-scaling of the times for the analysis as $\tau = t / t_{\rm KL, i=90\degree}$, hence averaging over KL cycles.
\cite{Katz_2011} show that the evolution of $\Omega_e$ and $\Theta$ are given by
\begin{align} 
    \dv{\Omega_e}{\tau} &= \Theta \qty(\frac{6 E\qty(x) -  3 K\qty(x)}{4 K\qty(x)}) \label{eq:omega_e} \ ,
    \\ \dv{\Theta}{\tau} &= \frac{-15 \pi \epsilon}{64 \sqrt{10}} \frac{\sqrt{\Theta} \sin{\qty(\Omega_e)}}{K\qty(x)} \qty(4 - 11 C_{\rm KL} \sqrt{6 + 4 C_{\rm KL}}) \label{eq:theta_evolution} \ ,
\end{align}
where $K\qty(x)$ and $E\qty(x)$ are complete elliptic functions of the first kind with $x\qty(C_{\rm KL}) = {3 \qty(1 - C_{\rm KL})}/({3 + 2 C_{\rm KL}})$.

Another constant of motion found by \cite{Katz_2011} is
\begin{align}
 \chi = F\qty(C_{\rm KL}) - \epsilon \cos{\qty(\Omega_e)} \ ,   
\end{align}
which connects the dynamics of $C_{\rm KL}$ with $\Omega_e$, and where
\begin{align}
    F\qty(C_{\rm KL}) = \frac{32 \sqrt{3}}{\pi} \int_{x\qty(C_{\rm KL})}^{1} \frac{K\qty(\eta) - 2 E\qty(\eta)}{\qty(41 \eta - 21) \sqrt{2 \eta + 3}} \dd{\eta} \ .
\end{align}
From Equation~\eqref{eq:phi_constant}, we can compute the evolution of $C_{\rm KL}$ as
\begin{align} \label{eq:CKL_evolution}
    \dv{C_{\rm KL}}{\tau} = - \frac{1}{2} \Theta \ ,
\end{align}
from which we can extract the period of the motion
\begin{align} \label{eq:tau_EKM}
    \tau_{\rm EKM} = \oint \dd{\tau} = \oint \frac{\dd{C_{\rm KL}}}{\dot{C}_{\rm KL}} \ .
\end{align}
Now, by combining Equation~\eqref{eq:tau_EKM} with Equation~\eqref{eq:CKL_evolution} and thanks to Eqs.~\eqref{eq:omega_e} and~\eqref{eq:theta_evolution}, \cite{Antognini_2015} has shown that the exact EKM period can be obtained as
\begin{align}
    \tau_{\rm EKM} =& \frac{256 \sqrt{10}}{15 \pi \epsilon} \int_{C_{\rm KL, min}}^{C_{\rm KL, max}} \frac{K\qty(x) \dd{C_{\rm KL}}}{\sqrt{2 \qty(\Phi_{\rm quad} - C_{\rm KL})} \qty(4 - 11 C_{\rm KL})} \nonumber
    \\ &\times \qty[\qty(1 - \frac{\chi - F\qty(C_{\rm KL})^2}{\epsilon}) \qty(6 + 4 C_{\rm KL})]^{-1/2} \ .
\end{align}
Finally, \cite{Antognini_2015} extracts the EKM timescale (see Equation~\ref{eq:tEKM}) as a function of the correct $\epsilon$ dependency and shows that it matches the EKM period (see Figure 5 from \citealp{Antognini_2015}).

\section{General relativity effects} \label{sec:appendix_GR}
General relativity (GR) effects, such as the periapsis precession can prevent Kozai-Lidov oscillations by stopping the Kozai resonance~\citep{Holman_1997}.
For the resonance to occur, the Kozai-Lidov timescale must be shorter than the GR precession timescale \citep{Blaes_2002}
\begin{align}
\frac{t_{\rm GR, inner}}{t_{\rm KL}} >1 \ ,
\end{align}
where the GR precession timescale is given by
\begin{align}
    t_{\rm GR} = \frac{1}{3} \frac{a_1}{c} \qty[\frac{a_1 c^2}{G\qty(m_0+m_1)}]^{3/2} \qty(1 - e_1^2) \ ,
\end{align}
following~\cite{Fabrycky_2007,Blaes_2002}. In the quadrupolar regime
\begin{align}
    \frac{t_{\rm GR, inner}}{t_{\rm KL}} &= \frac{16}{5} \frac{a_2^3}{a_1^4} \frac{\qty(1 - e_2^2)^{3/2}}{1 - e_1^2} \qty(\frac{G}{c})^{2} \frac{\qty(m_0 + m_1)^{3/2}}{m_2} \ ,
    \\ & \sim 1.6 \times 10^{5} \ \qty(\frac{a_2}{10 {\rm A.U.}})^3 \qty(\frac{a_1}{1 {\rm A.U.}})^{-4} \ ,
\end{align}
for circular orbits and, $m_0 = 1.4\, M_\odot$, $m_1 = 10^{-11}\, M_\odot$ and $m_2 = 10 \, M_\odot$. In the octupolar regime, combining Equation~\eqref{eq:tEKM}, we found
\begin{align}
    \frac{t_{\rm GR, inner}}{t_{\rm EKL}} &= \frac{128 \sqrt{10}}{15 \pi \sqrt{\epsilon}} \frac{t_{\rm KL}}{t_{\rm GR, inner}} \ , \nonumber
    \\ & \frac{2048 \sqrt{10}}{75 \pi \sqrt{\epsilon}} \frac{a_2^3}{a_1^4} \frac{\qty(1 - e_2^2)^{3/2}}{1 - e_1^2} \qty(\frac{G}{c})^{2} \frac{\qty(m_0 + m_1)^{3/2}}{m_2} 
    \\ & \sim 1.7 \times 10^{3} \ \qty(\frac{\epsilon}{0.01})^{-1/2} \qty(\frac{a_2}{10 {\rm A.U.}})^3 \qty(\frac{a_1}{1 {\rm A.U.}})^{-4}  \ ,
\end{align}
for elliptical orbits $e_1=e_2=0.1$ and, $m_0 = 1.4\, M_\odot$, $m_1 = 10^{-11}\, M_\odot$ and $m_2 = 10 \, M_\odot$. For both regimes, the GR precession effects are subdominant.

\begin{figure}[tb!]
	\center
	\includegraphics[width = 0.99\columnwidth]{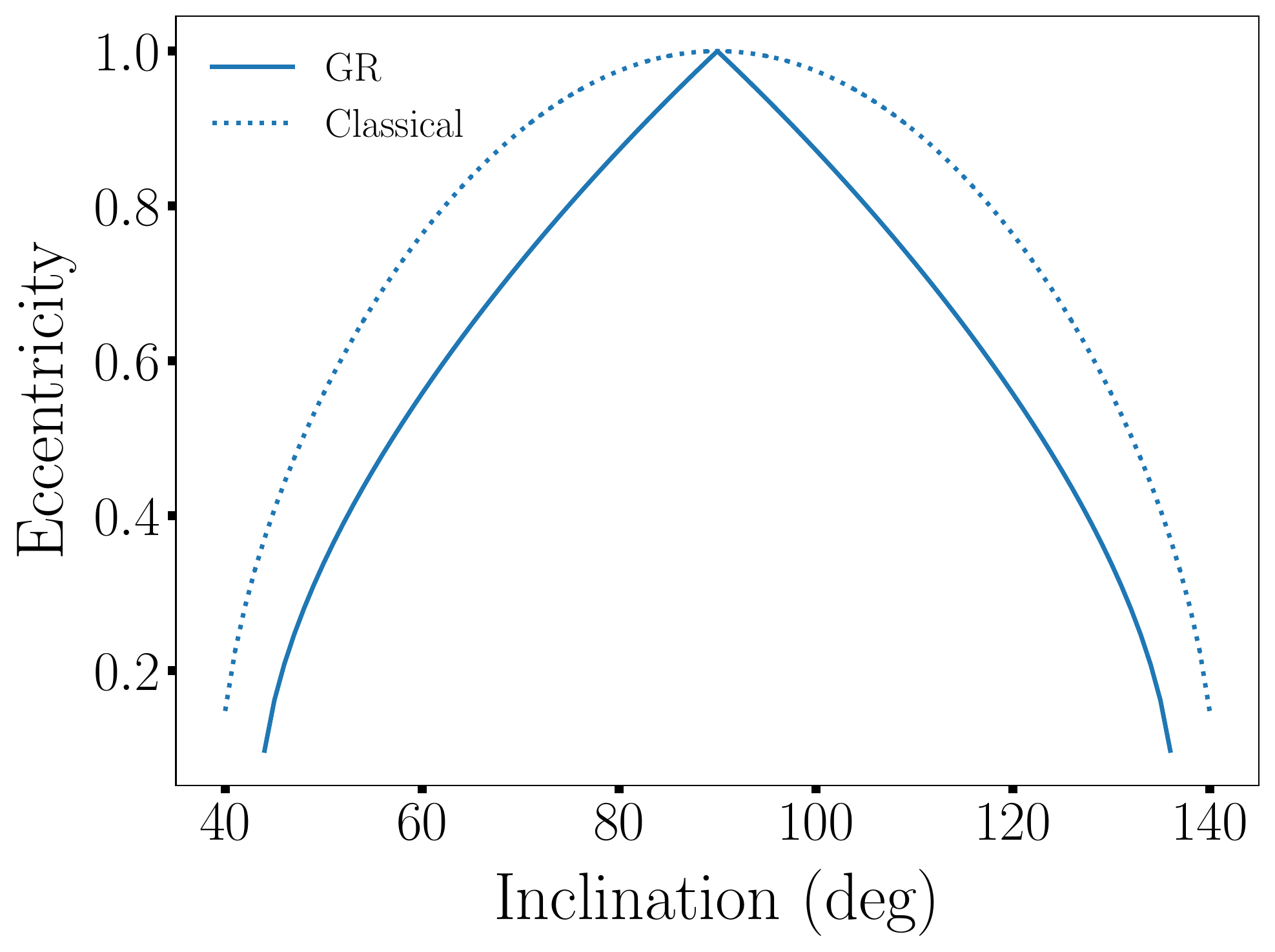}
	\caption{GR effects in the quadrupolar regime: maximal eccentricity reachable via Kozai-Lidov perturbations as a function of the inclination, for the classical computation (dotted line) and the general relativistic corrections (straight line). The GR corrections tends to reduce the maximal eccentricity reachable.}
	\label{fig:ecc}
\end{figure}

In the quadrupolar regime, the GR precession effects can suppress high eccentricity excitation $e_{\rm 1, max, KL}$ for the inner orbit. The ratio between the inner orbit GR precession timescale and the Kozai-Lidov timescale can be rewritten as \citep{Naoz_2016,2015MNRAS.447..747L}
\begin{align}
\frac{t_{\rm GR, inner}}{t_{\rm KL}} = \epsilon_{\rm GR}^{-1}(1-e_{\rm 1, max, GR}^2) \ ,
\end{align}
where
\begin{align}
\epsilon_{\rm GR} =\frac{3 G \qty(m_0+m_2)^2 a_{2}^3 (1 - e_2^2)^{3/2}}{{a_1^4 c^2 m_1}}\ .
\end{align}
The maximal eccentricity reachable $e_{\rm 1, max, GR}$ taking into account GR effects satisfies the following equation
\begin{align}
\epsilon_{\rm GR} \left(\frac{1}{j} - 1\right) = \frac{9}{8} \frac{e_{\rm 1, max, GR}^2}{j^2}\left[j^2 - \frac{5}{3}\cos^2{i}\right]\ ,
\end{align}
with $j = ({1 - e_{\rm 1, max, GR}^2})^{1/2}$.
For $\epsilon_{\rm GR} \ll 1$, this yields 
$j \approx ({{15^{1/2}}}/{3})\cos{i}  = ({{15}^{1/2}}/{5}) (1 - e_{\rm max, KL}^2)^{1/4}$. Figure~\ref{fig:ecc} displays both maximal eccentricities computed in the classical and general relativistic frameworks.
As demonstrated also in~\cite{2015MNRAS.447..747L}, GR effects on the maximum eccentricity reached will be stronger for higher inclinations.

\section{Quadrupolar treatment}\label{app:quadrupolar}
In this section, we focus on the quadrupolar regime, where we show that despite the dynamical differences between the two regimes, quadrupolar effects can also be efficient in driving asteroids close to the pulsar and produce FRB-like emissions on timescales compatible with the observed FRB rates.
However unlike the octupolar regime, the maximal eccentricity reachable is bounded by the initial inclination of the asteroid (since no flip occurs, the eccentricity does not increase to $i=90\degree$). This difference implies that not all asteroids triggering Kozai-Lidov oscillations are able to reach the Roche limit (closest position to the central pulsar, hence required to produce the strongest radio emissions). Therefore in the first part of this Appendix, we derive a criterion to discriminate between asteroids potentially able to produce radio emissions and those which are not. From this criterion, we then compute the fraction of asteroids inside a belt participating to the emissions. We also take into account GR corrections, due to precession effects, on the maximal eccentricity reachable. And finally, we present the results of the quadrupolar approach in this Kozai-Lidov induced FRB emissions.

\subsection{Roche limit crossing criteria} \label{app:roche_limit}
Under the influence of Kozai-Lidov oscillations, the eccenctricity of the inner binary can reach values sufficiently high so that its periastron crosses the Roche limit.
This limit represents the closest possible position for an object in orbit, beyond which it be disrupted by tidal forces.
The periastron of the elliptical orbit of the inner binary is given by
$r^{\rm periastron}_{\rm ast} = a_{\rm ast} \qty(1 - e_{\rm ast})$. 
If the periastron crosses the Roche limit (Equation~\ref{eq:Roche}), the following equation is verified:
\begin{align}
a_1\qty(1 - e_1) = 2 R_1 \left(\frac{m_0}{m_1}\right)^{1/3}\ ,
\end{align}
where $R_1$ and $R_2$ are the inner binary object radius.
Kozai-Lidov oscillations can drive the inner binary orbit down to the Roche limit if the following condition is fulfilled:
\begin{align}
\sqrt{1 - 5/3 \cos^2{i_{\rm tot}}} \geqslant 1 - 2 \frac{R_1}{a_1} \left(\frac{m_0}{m_1}\right)^{1/3}\ .
\end{align}
Namely, we require the maximal Kozai-Lidov eccentricity to be larger than the required eccentricity for the periapsis to cross the Roche limit.
Any inner binary, with orbital parameters matching the above equation, will be disrupted by tidal forces on a secular timescale following the Kozai-Lidov timescales.

\subsection{Fraction of asteroids crossing the Roche limit due to KL oscillations}\label{app:fraction_KL}
Assuming that the initial distribution of $a_{\rm ast}$ in the belt follows a Normal distribution with mean $\langle a_{\rm ast}\rangle$ and width $\sigma_a=\varepsilon_{\rm ast}\langle a_{\rm ast}\rangle$, the mean distance between two consecutively falling asteroids can be estimated statistically as $\langle\Delta a_{\rm ast}\rangle\approx \sigma_a/N_{\rm ast,KL}$, with $N_{\rm ast,KL}$ the number of asteroids meeting the Kozai-Lidov criterion. One can express $N_{\rm ast,KL}=f(i_c)\, N_{\rm ast}$, with $N_{\rm ast}$ the total number of asteroids in the belt and $f(i_c)$ the fraction of asteroids meeting the Kozai-Lidov criterion.

\begin{figure}[tb!]
	\center
	\includegraphics[width = 0.99\columnwidth]{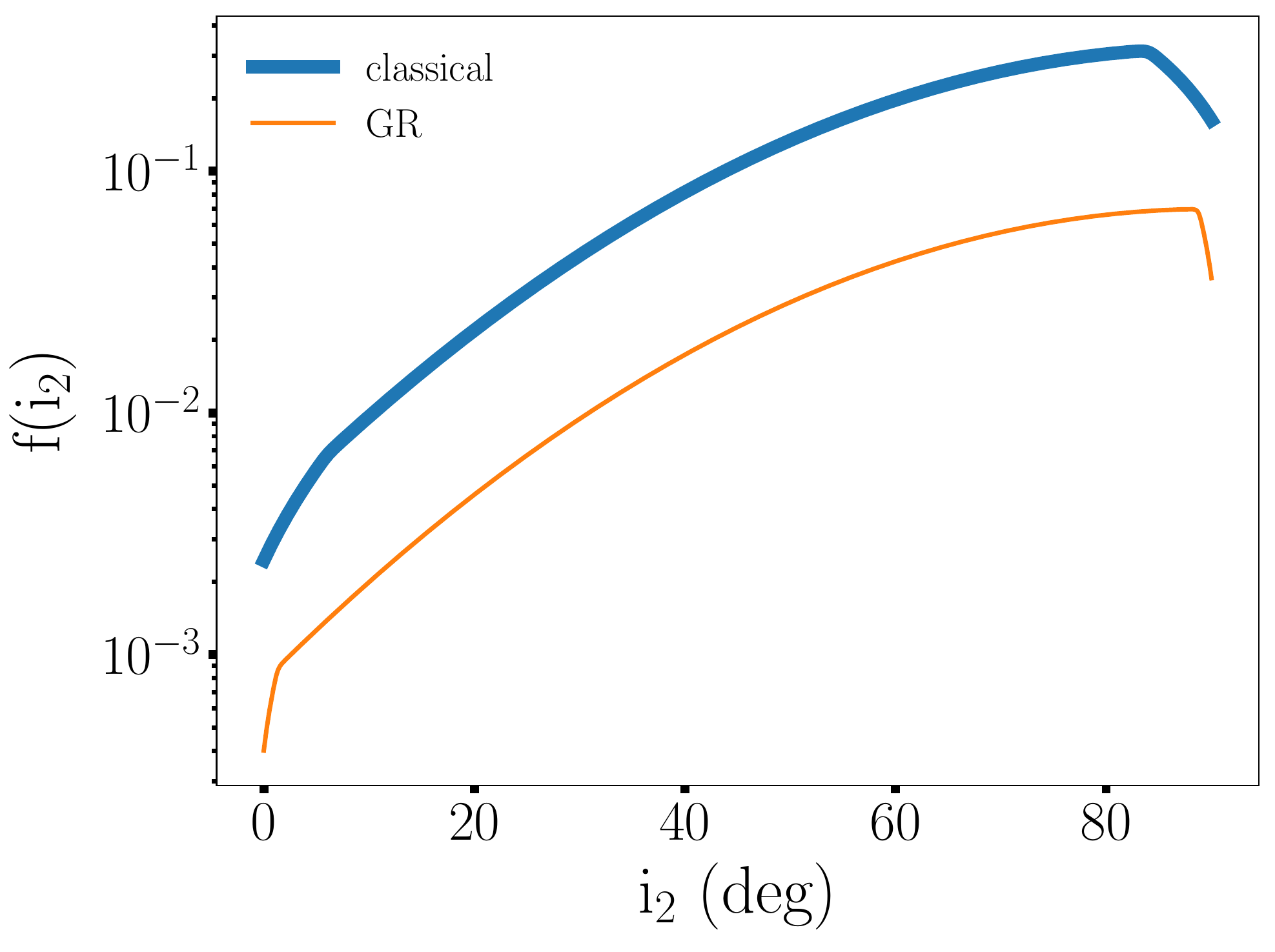}
	\caption{Fraction of asteroids reaching the Roche limit via the quadrupolar Kozai-Lidov effect in the classical calculation (blue) and in the general relativity case (orange), as a function of the outer body inclination $i_2$. Its mass is set to $m_1=10\,M_\odot$. The density of the asteroids are set to $\rho_{\rm ast}=2\,$g.\,cm$^{-3}$ and their semi-major axes follow a Normal law with mean semi-major axis $\langle a_{\rm ast} \rangle =1\,{\rm A.U.}$ and standard deviation $\sigma_{a} = 0.15\, \langle a_{\rm ast} \rangle$.}
	\label{fig:fi2}
\end{figure}

The fraction $f(i_c)$ of asteroids meeting the Kozai-Lidov criterion depends on the inclination $i_c$ as
\begin{align}\label{eq:fi2}
f(i_c) = \int_{i_{\rm ast}=0}^{\pi / 2}\!\!\! \mathcal{N}_{\langle i_{\rm ast}\rangle}^{\sigma_{i}}\int_{a_{\rm ast}=0}^{a_{\rm ast, KL}}\!\!\!\!\!\!\!\!\!\!\!\!\mathcal{N}_{\langle a_{\rm ast}\rangle}^{\sigma_{a}}(a_{\rm ast}) {{\rm d}a_{\rm ast}\,{\rm d}i_{\rm ast}}\ ,
\end{align}
where $\mathcal{N}_{\langle x\rangle}^{\sigma_x}(x)$ is the normal distribution function of mean $\langle x\rangle$ and variance $\sigma_x^2$. The Kozai-Lidov maximum semi-major axis to reach the Roche limit reads 
\begin{align}
a_{\rm ast,KL}(i_c)=\frac{2R_{\rm ast}\left({M_{\rm NS}}/{M_{\rm ast}}\right)^{1/3}}{1-\left[1-({5}/{3})\cos^2\left(i_{\rm ast}+i_c\right)\right]^{-1/2}}  \ .
\end{align}
Figure~\ref{fig:fi2} presents the values of $f(i_c)$ in the classical derivation (blue). However, we will see in the next paragraph that in our regime, General Relativity (GR) effects dominate and lead to lower $f(i_c)$. 

\subsection{General Relativity corrections}
As shown in Appendix~\ref{sec:appendix_GR}, the GR corrections tend to reduce the maximal eccentricity reachable and therefore translates to a lower fraction of asteroids capable of reaching the Roche limit.

The Kozai-Lidov oscillations in the GR regime can drive an asteroid to  disruption inside the Roche limit if the following condition is fulfilled 
\begin{align} \label{eq:eta}
&a_{\rm ast, KL, GR}(i_c) =\qty[2 R_{\rm ast} \qty(M_{\rm NS}/M_{\rm ast})^{1/3}] \nonumber
\\ &\times \qty[1 - \sqrt{1 - 16/25 \sqrt{(5/3) \cos^2(i_{\rm ast} + i_c)}}]^{-1} \ .
\end{align} 
This maximum semi-major axis replaces $a_{\rm ast,KL}$ in Equation~(\ref{eq:fi2}), leading to a reduction of $f(i_c)$ by a factor $\sim3$, as can be seen in Figure~\ref{fig:fi2}. Numerically, including GR corrections, $f(i_c=45\degree)\sim 0.02$, leading to $N_{\rm ast,KL}=f(i_c=45\degree)(N_{\rm ast}/1000)\sim 20$. 

In this calculation, we have neglected the tidal and rotation terms, which can also affect the maximum eccentricity reached by the body. These terms are negligible compared to the GR term in our model. We note that the quadrupole approximation leads to a good analytical estimate of the orbital evolution, even when the octupole effects are strong~\citep{2015MNRAS.447..747L}. \\

\subsection{Results}
Thanks to the framework detailed in Appendices~\ref{app:fraction_KL},~\ref{sec:appendix_GR} and~\ref{app:roche_limit} we can compute the infall rates of asteroids in binary pulsar systems, in the quadrupolar regime of the Kozai-Lidov effect.
The mean relative Kozai-Lidov time between two consecutive asteroid disruptions can then be estimated as
\begin{align}\label{eq:mean_dtKL}
    \langle\Delta t_{\rm KL} \rangle &\sim  150\,{\rm yr}\,\left[\frac{f(i_c)N_{\rm ast}}{20}\right]^{-1}\frac{\varepsilon_{\rm ast}}{0.15}\left(\frac{\langle a_{\rm ast}\rangle}{{\rm A.U.}}\right)^{-3/2} \nonumber\\
    &\times \left(\frac{a_c}{100\,{\rm A.U.}}\right)^3\left(\frac{M_c}{10\,M_\odot}\right)^{-1}\left(\frac{M_{\rm NS}}{1.4\,M_\odot}\right)^{1/2}\ ,
\end{align}
where we have assumed $e_c=0$ for the numerical estimate. Here, we have used the parameter values of typical NSBH/NSMS systems. 
The value of $\varepsilon_{\rm ast}$ is chosen so as to fit the parameters of the Solar belt (see Section~\ref{section:application_belt}).

The factor $N_{\rm ast,KL}=f(i_c)N_{\rm ast}$ corresponds to the number of asteroids which experience the Kozai-Lidov effect. 
Higher $i_c$ can boost the FRB rates estimated above by $1-2$ orders of magnitude, due to a larger $f(i_c)$. Larger $N_{\rm ast}$ and higher $i_c$ would also shorten $\langle\Delta t_{\rm KL}\rangle$, consequently increasing the event rates of Section~\ref{section:FRBrates}.

\begin{figure}[tb!]
	\center
	\includegraphics[width = 0.99\columnwidth]{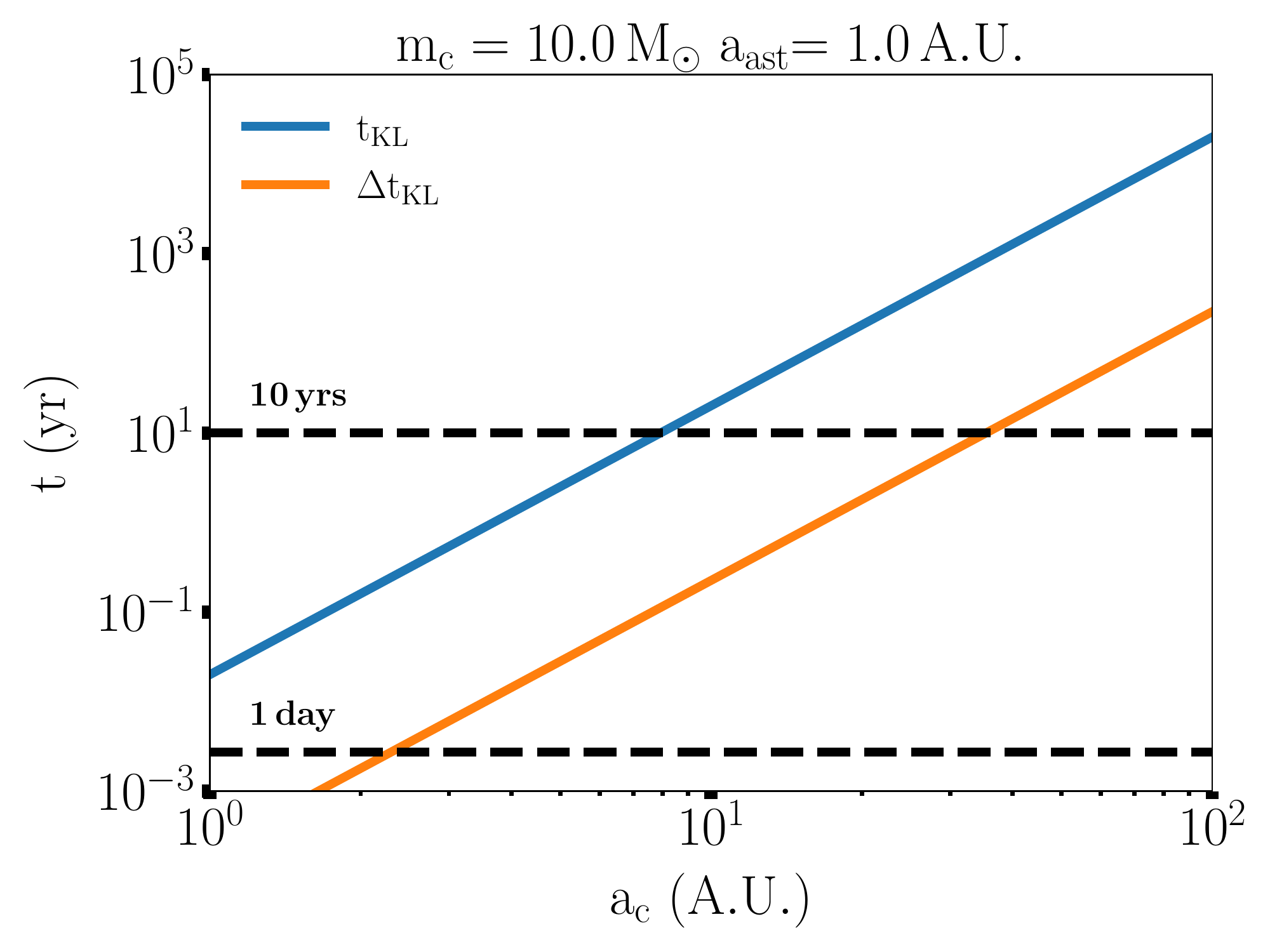}
	\caption{Mean Kozai-Lidov time $\expval{t_{\rm KL}}$ (Equation~\ref{eq:tKL}) and relative time delay $\expval{\Delta t_{\rm KL}}$ (Equation~\ref{eq:mean_dtKL}) as a function semi-major axis $a_c$ of the outer perturbing body semi-major axis.}
	\label{fig:quad_KL_times}
\end{figure}

Figure~\ref{fig:quad_KL_times} displays the evolution of the Kozai-Lidov time $t_{\rm KL}$ and relative delay $\Delta t_{\rm KL}$ as a function of the outer perturbing body semi major axis $a_{\rm c}$ (here $a_{\rm BH}$). A clear distinction between close and wide systems can be made based on the typical delay, the first ones, present relative delay on day-scales while the other ones have relative delay of several tens of years.

\begin{figure*}[tb!]
	\center
	\includegraphics[width = 0.49\linewidth]{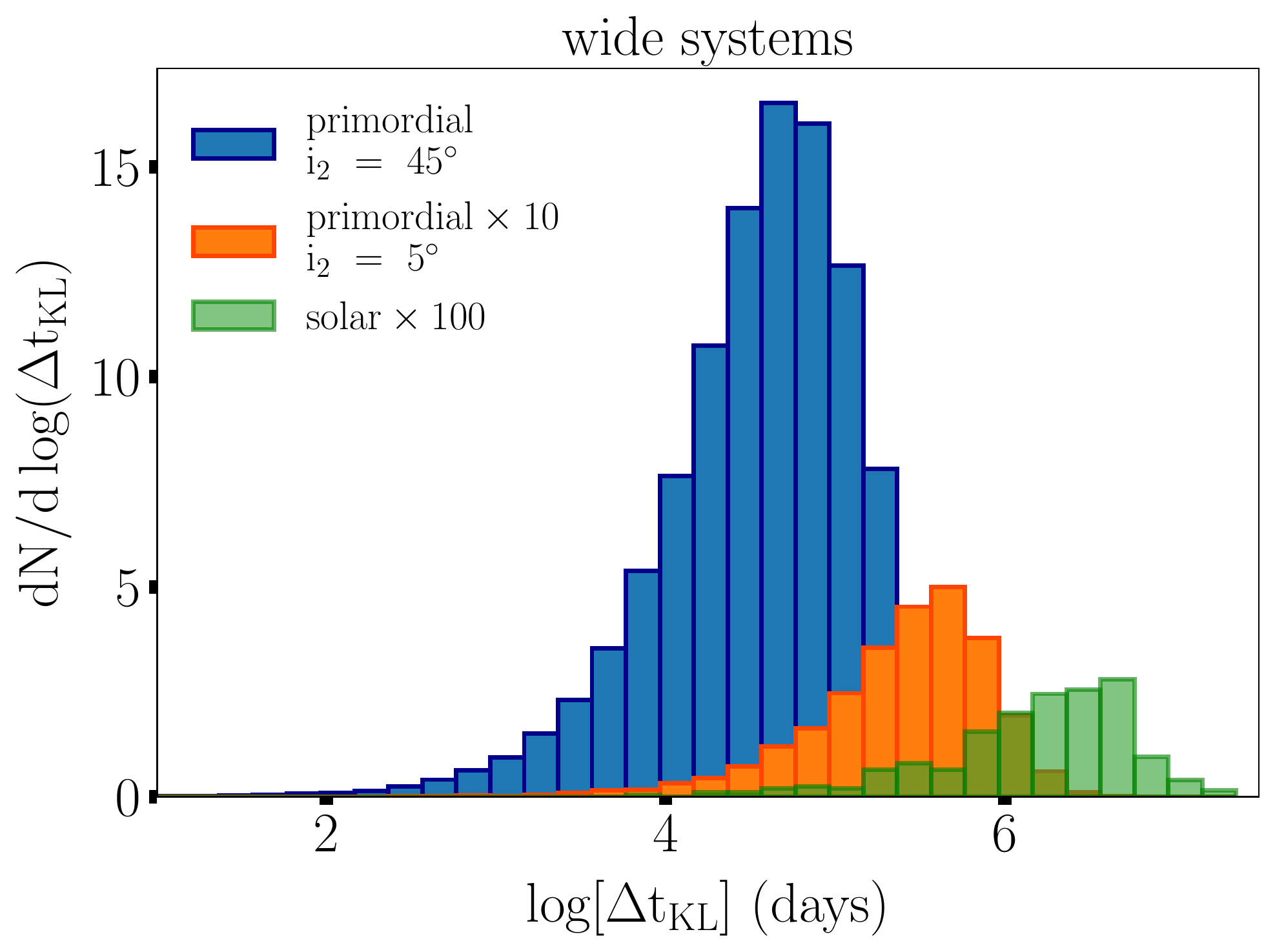}
	\includegraphics[width = 0.49\linewidth]{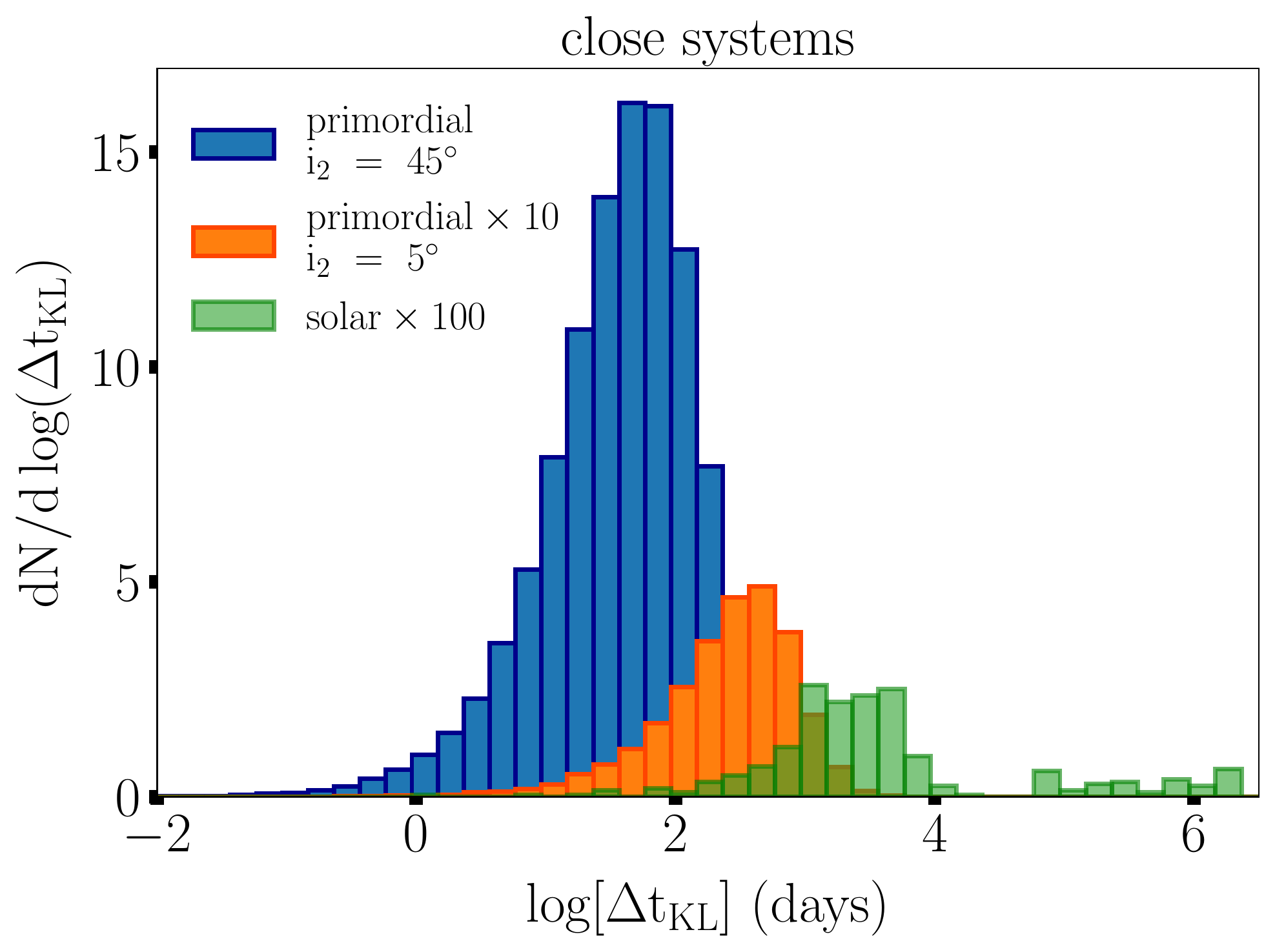}
	\caption{Same as Figures~\ref{fig:rates_inclinations} and~\ref{fig:rates_light_heavy}, distribution of relative time differences $\Delta t_{\rm KL}$ of falling asteroids in the Roche lobe of the central compact object due to quadrupolar Kozai-Lidov oscillations, for the current Solar asteroid belt (green) and the primordial belt, for an inclination of the outer body hole plane $i_c=5\degree$ (orange) and $i_c=45\degree$ (blue), and initial asteroid number $N_{\rm ast}=10^3$. We consider a high mass $M_c=10\, M\odot$ wide system ({\it left}) with $a_c = 100$\,A.U. and $\expval{a_{\rm ast}}= 1$\,A.U., and a close system ({\it right}) with $a_c = 10$\,A.U. and $\expval{a_{\rm ast}}= 1$\,A.U.}
	\label{fig:KL_rate}
\end{figure*}

These analytical results are confirmed thanks to the simulations detailed in Section~\ref{section:application_belt}, and here used only in the quadrupolar regime.
Figure \ref{fig:KL_rate} shows the distribution of the relative time delays for asteroids falling onto the central neutron star, in the quadrupolar regime, for a current (green) and primordial Solar-like belt for companion inclinations $i_{\rm c}=5\degree$ (orange) and $i_{\rm c}=45\degree$ (blue). The central neutron star has mass $M_{\rm NS} = 1.4\,M_{\odot}$ and the outer companion $M_{\rm c} = 10\,M_{\odot}$. The initial number of asteroids is set to $N_{\rm ast}=10^3$.
We examine the case of a wide system with companion distance $a_{\rm c}=10$\,A.U. and mean asteroid belt distance $\expval{a_{\rm ast}}=1$\,A.U. (left panel).

For the wide system, the infall rates span days to millions of years, with a maximum around $\langle \Delta t_{\rm KL} \rangle\sim 10s-100s$\,years, depending on the inclination $i_c$. For close systems, the rates are of order day-scales.

The comparison between the current Solar belt and the primordial belt shows that the lack of Kirkwood gaps induces a drastic increase of short time-scales in the asteroid infall rate, and depending on the inclination, a factor of a few to an order of magnitude more events in total. Larger inclinations $i_c$ lead to shorter time-scales, and to higher event rates since the shifting time-scales  due to inclination $i_c$ is dominant over the $1/f(i_c)$ effect. Systems with larger inclinations $i_c$ and with higher rates ($\Delta t_{\rm KL}\sim 10\,$yr) will thus dominate in the sky. 

Furthermore, one can notice the tail distribution at large time-scales for the Solar belt in the right panel of Figure~\ref{fig:KL_rate}. It  results from the Kirkwood gaps, where groups of asteroids with lower inclinations can reach the Roche limit due to the closer position of the outer black hole $10\,$AU (rather than $100\,$AU in the left panel). However, their lower inclinations result in larger time-scales.

\begin{figure}[tb!]
	\center
	\includegraphics[width = 0.99\columnwidth]{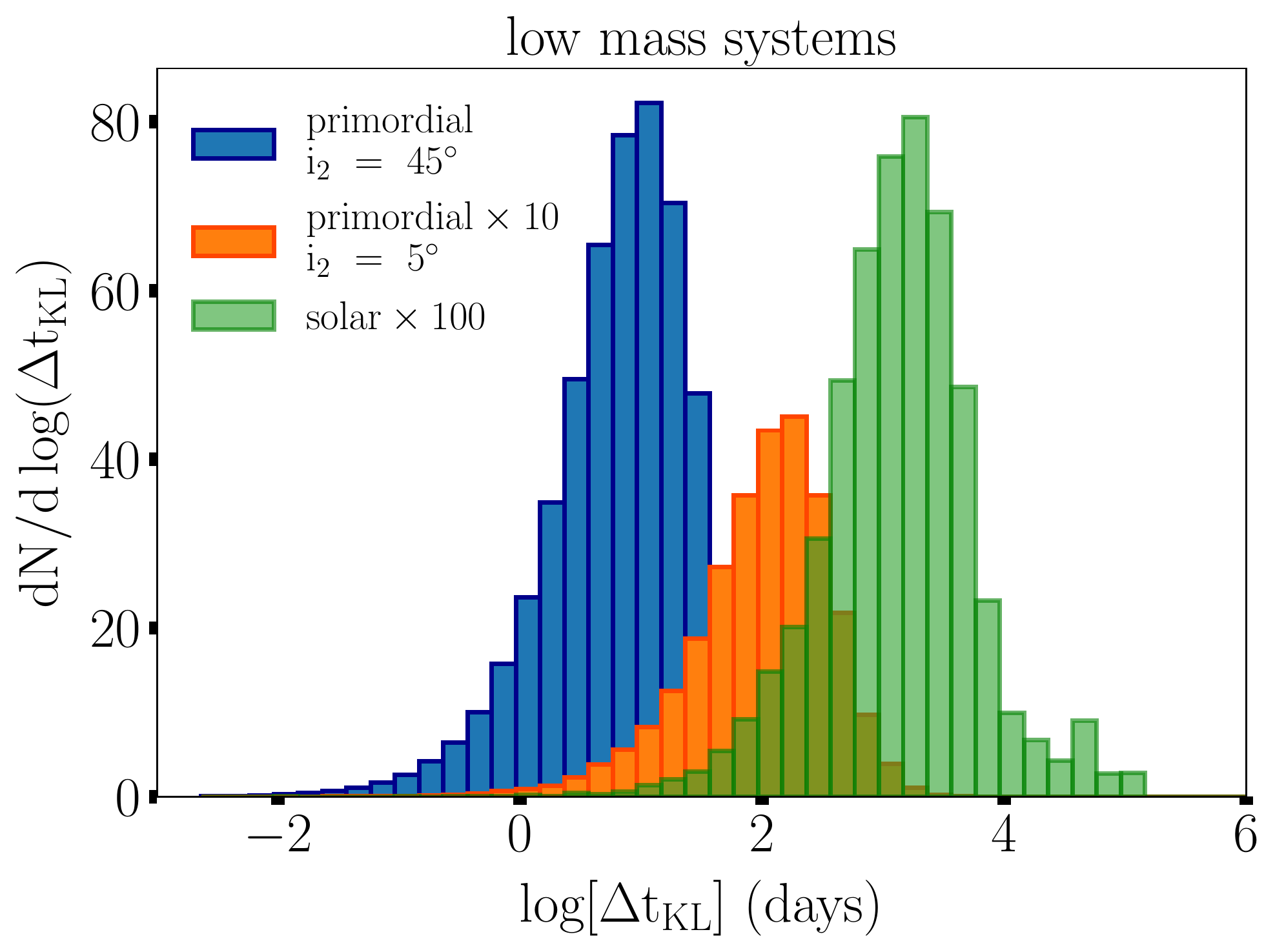}	
	\caption{Same as Figure~\ref{fig:KL_rate}, but for a low mass $M_c = 1\,M_{\odot}$ close system $a_c=2\,$A.U. and $\langle a_{\rm ast}\rangle= 0.2\,$A.U.  The initial asteroid number is $N_{\rm ast}=10^3$.}
	\label{fig:KL_rate_compact}
\end{figure}

Figure~\ref{fig:KL_rate_compact} shows a consistent result with Figure~\ref{fig:KL_rate}, a low mass companion in a more close system induces short timescales. The reduction of the Kozai-Lidov effect due to the low mass companion is compensated with shorter distances in that system, ending up with a similar result than in the left panel of Figure~\ref{fig:KL_rate} (high mass companion in close system).
Finally the result obtained for the quadrupolar treatment are fully consistent with the octupolar approach, in the sense that the dichotomy between repeaters and non-repeaters is explained thanks to the populatin of systems involved in the Kozai-Lidov mechanism: either close systems leading preferentially to repeater FRBs or wide systems leading preferentially to non-repeater FRBs.

\end{document}